\documentclass{ieeeaccess}

\usepackage{cite}
\usepackage{amsmath,amssymb,amsfonts}
\usepackage{algorithmic}
\usepackage{graphicx}
\usepackage{textcomp}
\usepackage{mathtools}
\usepackage{blindtext}
\usepackage{soul} 
\usepackage{booktabs} 

\def\mathlette#1#2{{\mathchoice{\mbox{#1$\displaystyle #2$}}%
                               {\mbox{#1$\textstyle #2$}}%
                               {\mbox{#1$\scriptstyle #2$}}%
                               {\mbox{#1$\scriptscriptstyle #2$}}}}
\renewcommand{\Vec}[1]{\mathlette{\boldmath}{#1}}

\newcommand{\be}{\begin{equation}}
\newcommand{\ee}{\end{equation}}
\def\oT{\ensuremath{^\text{T}}} 
\def\oH{\ensuremath{^\text{H}}} 
\def\diag{\text{diag}}

\def\BibTeX{{\rm B\kern-.05em{\sc i\kern-.025em b}\kern-.08em
    T\kern-.1667em\lower.7ex\hbox{E}\kern-.125emX}}
\begin{document}
\history{Date of publication xxxx 00, 0000, date of current version xxxx 00, 0000.}
\doi{10.1109/ACCESS.2017.DOI}

\bibliographystyle{IEEEtran}

\title{Multi-Node Vehicular Wireless Channels: Measurements, Large Vehicle Modelling, and Hardware-in-the-Loop Evaluation}

\author{\uppercase{Stefan Zelenbaba}
	\IEEEmembership{Member, IEEE} ,
	\uppercase{Benjamin Rainer},
	\uppercase{Markus Hofer} \IEEEmembership{Member, IEEE}, 
	\uppercase{David Löschenbrand}
	\IEEEmembership{Member, IEEE} ,
	\uppercase{Anja DakiĆ} \IEEEmembership{Graduate Student Member, IEEE} ,
	\uppercase{Laura Bernadó} \IEEEmembership{Member, IEEE},
	\uppercase{Thomas Zemen}
	\IEEEmembership{Senior Member, IEEE} }
\address{AIT Austrian Institute of Technology GmbH, 1210 Vienna, Austria}
\tfootnote{This paper is a result of the projects SCOTT (www.scott-project.eu), REALISM and DEDICATE. The REALISM project (864188) is funded by the Austrian Research Promotion Agency (FFG) and the Austrian Ministry for Climate Protection, Energy, Transport, Innovation and Technology (BMK) of the funding program transnational projects. The DEDICATE project is funded within the Principal Scientist grant “Dependable Wireless 6G Communication Systems” at the AIT Austrian Institute of Technology. The SCOTT project has received funding from the Electronic Component Systems for European Leadership Joint Undertaking under grant agreement No 737422. This Joint Undertaking receives support from the European Union’s Horizon 2020 research and innovation programme and Austria, Spain, Finland, Ireland, Sweden, Germany, Poland, Portugal, Netherlands, Belgium, and Norway. The document reflects only the author’s view and the Commission is not responsible for any use that may be made of the information it contains.}

\markboth
{Zelenbaba \headeretal:Multi-Node Vehicular Wireless Channels: Measurements, Large Vehicle Modelling, and Hardware-in-the-Loop Evaluation}
{Zelenbaba \headeretal:Multi-Node Vehicular Wireless Channels: Measurements, Large Vehicle Modelling, and Hardware-in-the-Loop Evaluation}

\corresp{Corresponding author: Stefan Zelenbaba (stefan.zelenbaba@ait.ac.at)}

\begin{abstract} {Understanding multi-node vehicular wireless communication channels is crucial for future time-sensitive safety applications for human-piloted as well as partly autonomous vehicles on roads, railways and in the air. These highly dynamic wireless communication channels are characterized by rapidly changing channel statistics. In this paper we present the first fully mobile multi-node vehicular wireless channel sounding system, which is capable of simultaneously capturing multiple channel frequency responses, ensuring that measurement conditions are identical for all observed links. We use it to analyze road scenarios with multiple vehicles and a large obstructing double-decker bus.
The empirical measurement data is used to parametrize a model for the large obstructing vehicle within a geometry-based stochastic channel model. We compare the time-variant channel statistics obtained from our channel model with the ones from the measurement campaign. By means of a channel emulator we obtain the packet error rates of commercial modems for the measured and the simulated wireless communication channels and compare them, in order to validate the model at the link level. We find that the path loss, the root mean square (RMS) delay spread, and the RMS Doppler spread deviate by less than $3.6\,\mathrm{dB}$, $78\,\mathrm{ns}$, and $52\,\mathrm{Hz}$, respectively, for $80\%$ of the total simulation duration. The PER obtained from measured data is within the maximum and minimum bounds of our model for $86\%$ of the simulation duration.
}
 
\end{abstract}


\titlepgskip=-15pt

\maketitle

\section{Introduction}
\label{sec:introduction}

\PARstart{R}{eliable} and low-latency communication systems are crucial to connect multiple vehicles in challenging traffic scenarios. This allows for increased road safety and traffic efficiency by complementing sensors limited to line-of-sight (LOS), such as video or LIDAR (light detection and ranging), with sensor data from other vehicles in the vicinity \cite{Chen2017, 5g2016case}, as well as the exchange of kinematic information. 

Therefore, multi-node vehicular wireless channels need to be thoroughly analyzed by gathering measurement data in realistic scenarios. The empirical data gathered can be used to obtain realistic channel models which are a key component for testing the reliability of vehicular communication systems in lab environments. 

Multi-node channel measurements allow the recording of time-variant frequency responses of multiple wireless links simultaneously. Measuring multiple channels at once is particularly useful for vehicular scenarios due to the non-stationary statistical properties of vehicular wireless communication channels, where sequentially recreating the exact measurement conditions in order to compare links between different nodes can be almost impossible.

Large-vehicles like buses\footnote{According to \cite{Eurostat_transport2020} buses accounted for $9.4\%$ of the inland passenger traffic in the European Union in 2017.} can block the line-of-sight between other vehicles. Therefore, it is essential to include the detailed characteristics of this vehicle category in a channel model to better reproduce realistic traffic situations. Furthermore, the correct choice of a channel model is crucial to efficiently, yet accurately, capture the statistics of the time- and frequency-variant fading process in highly dynamic multipath propagation environments. This enables the design of reliable and affordable test procedures on the link and system levels.

Geometry-based stochastic channel models (GSCMs) present a trade-off between low-complexity stochastic channel models and the accuracy of ray-tracing channel models \cite{Almers_07_channelmodels}. We aim to achieve a high level of model accuracy while maintaining low model complexity and automating geometry modelling efforts, which makes GSCM a suitable choice.

In this work we therefore encompass the whole process. We present a multi-node measurement system, analyse collected measurement data and use it to calibrate a channel model. We then use both the measurement data and the model simulations to emulate a link between two commercial modems in our lab and compare their link-level packet error rate (PER).

\subsection{Contributions}

The contributions of this paper can be summarized as follows:
\begin{itemize}
	\item To the best of the authors' knowledge, in this paper we present the first vehicular \emph{multi-node} channel sounding system and measurement data from vehicular scenarios.
	\item Our results show the impact of a double-decker bus on link attenuation, root mean square (RMS) delay, and RMS Doppler spread in an urban overtaking scenario and an intersection scenario. 
	\item We present a GSCM that obtains its geometry data from the OpenStreetMap (OSM) \cite{OSM} database and includes a simple, yet accurate, model for a double-decker bus, calibrated by measurement data.
	\item We obtain time-variant PERs from measured and simulated channel transfer functions, by emulating a link between two commercial vehicular modems. The evaluated PERs are compared to validate our model at the link-level.
\end{itemize}  

\subsection{Organization of the paper}
The rest of this paper is organized as follows.
Section~\ref{sec:related_work} is dedicated to the discussion of related work.
In Section~\ref{sec:sounder} we present the concept of our multi-node channel sounder and the sounding principle.
Section~\ref{sec:meas_campaign} presents the measurement campaign setup, measurement scenarios and measurement results, with a discussion on large vehicle influence.
The channel model is presented and its results are compared with measurements in terms of time-variant channel statistics in Section~\ref{sec:gscm}.
We verify our model through link-level emulation in Section~\ref{sec:per_emulation}.

\subsection{Related Work}
\label{sec:related_work}
The benefits of multi-node channel sounding are highlighted in \cite{Wassie19}, where the authors present a multi-node channel sounder whose centralized synchronization limits its use with respect to dynamic scenarios. A combination of two multiple-input multiple-output (MIMO) channel sounders is used in \cite{Almers08} to capture two links simultaneously in indoor multi-user MIMO scenarios. Similar scenarios are investigated in \cite{Bauch07} and \cite{Bui13}. However, the measurement systems used in \cite{Almers08}, \cite{Bauch07} and \cite{Bui13} do not cover links between all nodes and are therefore unsuitable for measuring multi-node (or ad-hoc) communication scenarios.

A comparison of key features of the sounder used in this paper with previously developed multi-node channel sounding solutions is shown in Table~\ref{Multi_node_sounders}. While all three multi-node channel sounders that are listed operate below 6GHz, the AIT multi-node channel sounder is fully mobile, scalable to multiple nodes, and offers the largest measurement bandwidth ($150.25\,\mathrm{MHz}$).

\begin{table*}[hbtp]
	\begin{center}
		\caption{Existing multi-node channel sounding systems}
		\begin{tabular}{l l l l l}
        \toprule			
        Measurement system & Measurement Bandwidth & Frequency range & Scalability & Mobility\\
        \midrule
			AIT multi-node channel sounder \cite{Zelenbaba20_1} & 150.25 MHz         & <6 GHz               & yes                   & full      \\     
			\cite{Wassie19}                & 40 MHz          & <6 GHz               & yes                   & static          \\
			\cite{Almers08}                & $120\,\mathrm{MHz}$         & <6 GHz               & max. two channels                  & limited by cable length         \\
		\bottomrule
		\end{tabular}
		\label{Multi_node_sounders}
	\end{center}
\end{table*}

Simultaneous measurements of links between multiple vehicles are conducted in \cite{Nilsson15} and \cite{Nilsson18}, and the authors of \cite{Nilsson18} measure links between multiple vehicle groups that include trucks. However, both \cite{Nilsson15} and \cite{Nilsson18} only analyze the received power and measure PERs without inspecting time-variant channel impulse responses, which provide the ground truth and essential insights into the propagation properties of vehicular scenarios. In this paper we use a custom-built multi-node channel sounder to simultaneously collect the time-variant frequency responses between three vehicular nodes.

One of the initial efforts towards understanding vehicular channels in situations that include large vehicle obstructions is described in \cite{Gallagher06}, where the authors investigate the PER and communication range of a vehicular link obstructed by a truck. By measuring a static link between parked vehicles, the authors of \cite{He14} show that a school bus can introduce between $15\,\mathrm{dB}$ and $20\,\mathrm{dB}$ of attenuation between two vehicles, depending on the distance, and an increase of $100\,\mathrm{ns}$ in the RMS delay spread.

A channel sounding measurement with a truck between two cars in a dynamic environment is presented in \cite{Vlastaras14}, where the authors show the shadowing loss for different antenna arrangements and an increase in RMS delay spread when the truck is present. The average channel gain and RMS delay spread of vehicular channels obstructed by different types of vehicles and for different antenna positions, are investigated in \cite{Mahler16} for a rural overtaking scenario. The authors of \cite{Yang_20} measure a dynamic vehicular link with a bus between two cars and present a cluster-based channel model. Nevertheless, detailed channel sounding measurements that can be used to analyze and model the time-variant statistics of multi-vehicle scenarios that include large vehicles have not yet been performed.

The authors of \cite{Vlastaras17} describe a model for power contributions due to diffraction around a truck. However, different types of large vehicles have different structure, which needs to be considered in an attempt to make channel models more accurate.

We base the GSCM in this paper on the model presented in \cite{Karedal09}, which uses a simplified ray-tracing approach with a stochastic point scatterer distribution. However, the MIMO GSCM presented in \cite{Karedal09} is limited to vehicle-to vehicle communications in rural motorway and highway environments. The authors of \cite{Gustafson20} provide a generalized GSCM for urban intersections and they use high resolution measurements to identify scattering intersection points. However, although the authors of \cite{Gustafson20} use a general model, it shows moderate deviations when compared with measurement data. It should be noted that neither of the models in \cite{Karedal09} and \cite{Gustafson20} give attention to the impact of large vehicles. 

\section{Vehicular Multi-Node Channel Sounder}
\label{sec:sounder}

To collect time-variant frequency responses between all nodes in a single snapshot we use a custom-built software-defined radio (SDR) based multi-node channel sounding system \cite{Zelenbaba20_1} that exploits a scalable sounding scheme depicted in Fig.~\ref{fig:multi_sounder}. 

The scalable sounding scheme is a key component of the multi-node system and it uses the switched-array principle to measure $\binom{L}{2}=\frac{L!}{2(L-2)!}$ channels between $L$ nodes. Each snapshot interval $T_\mathrm{sys}$ is split into $L-1$ sounding intervals $T_\mathrm{s}$, or phases in which different links are measured. In case of an $L=3$ node system we split each snapshot into into two phases, where the second node switches from receiver (Rx) to transmitter (Tx) for the second phase as seen on the left side of Fig.~\ref{fig:multi_sounder}. 

\Figure[!ht]()[width=0.99\columnwidth, height=2in]{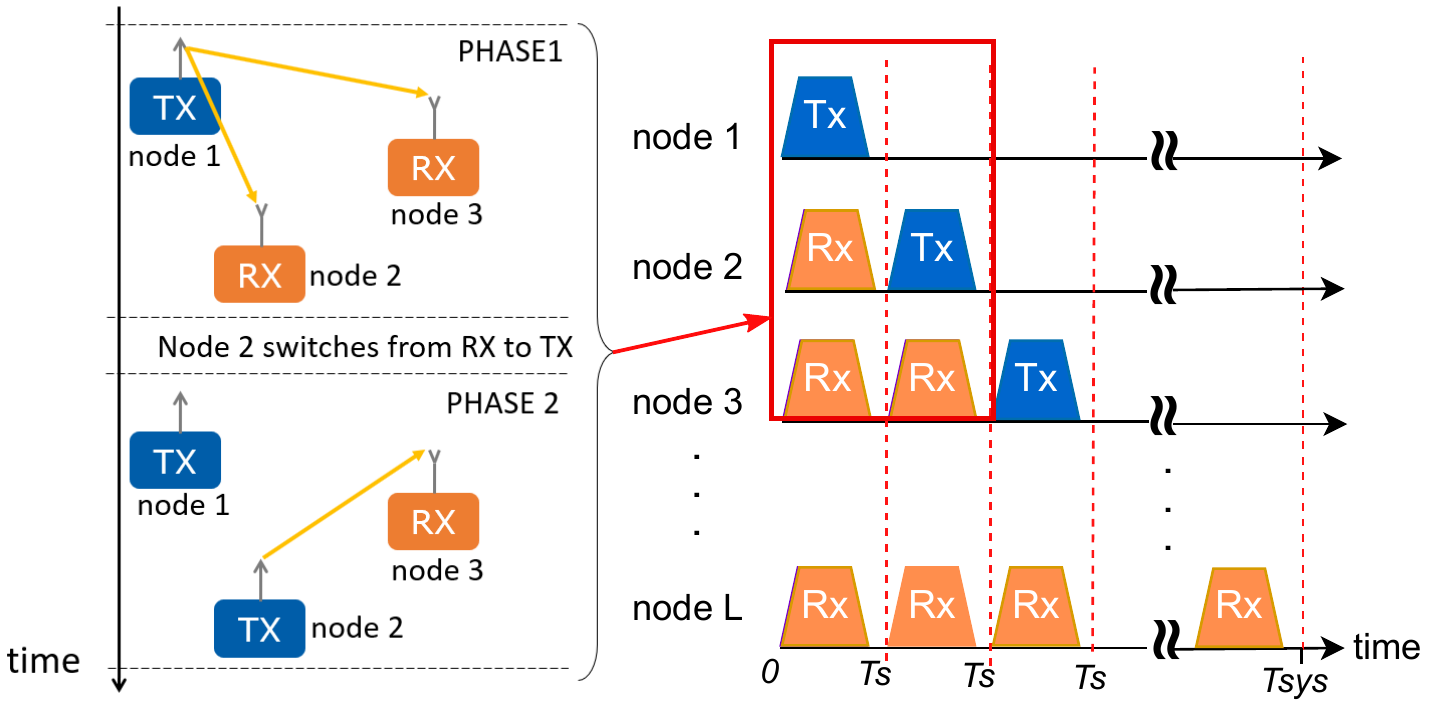}
{Sounding scheme for three nodes (left) and scalable sounding schedule for $L$ nodes (right). Measurement phases are delimited with dashed lines. During each phase, depending on the sounding schedule, a node can act either as a transmitter (Tx) or as a receiver (Rx). \label{fig:multi_sounder}}

The maximum resolvable Doppler shift of the multi-node sounder relates to the sounding interval as
\be
f_{\mathrm{Dmax}}=1/(2(L-1)T_\mathrm{s})\,.
\ee

By synchronizing all the nodes to a one pulse-per-second (1PPS) signal and choosing $T_{\mathrm{sys}}$ to be an integer fraction of the 1PPS, we align the 1PPS signal with the start of a snapshot. The operation of each node starts with the first 1PPS signal that the SDR receives from the Rubidium clock. Since the 1PPS signals of all nodes are previously synchronized, the snapshot synchronization on system level is ensured. This eliminates the need for starting trigger synchronization mechanisms and allows all nodes to operate on demand and move independently without range restrictions, enabling full mobility of the system and allowing measurements of complex multi-vehicle scenarios.

\subsection{Implementation}
The schematic of a single node of the multi-node sounder is shown in Fig.~\ref{fig:multi_scheme}. The National Instruments (NI) universal software radio peripheral (USRP) \cite{USRP} is used as radio frequency (RF) front-end. The USRP is equipped with a  field-programmable gate array (FPGA) chip that is used for raw data processing, RF chain selection and automatic gain control.

We use separate RF chains to transmit and receive, with an external amplifier added to the transmit chain and a PIN-diode switch placed before the shared antenna.
The system uses previously synchronized Rubidium clocks for the $10\,\mathrm{MHz}$ frequency reference signal and a one pulse-per-second (1PPS) signal for node synchronization.

\Figure[!ht]()[width=0.8\columnwidth]{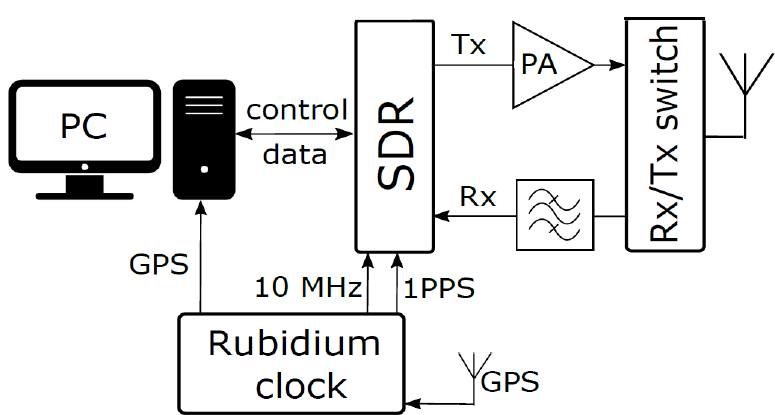}
{Schematic of a single AIT multi-node sounder node.
    The transmit (Tx) chain with a power amplifier (PA) and the receive (Rx) chain access the common antenna through a PIN-diode switch. The software defined radio (SDR) unit is controlled through the host PC, and it receives the reference signals from the Rubidium clock. \label{fig:multi_scheme}}

 Each sounder node is operated and configured through the LabView Communications software suite on a host PC where the recorded data is stored. Real-time calculation and display of the measured frequency responses allow live visual evaluation. 

\subsection{Sounding Principle}
The transmit signal $x[q]$ is designed using the algorithm from \cite{Friese97} in order to minimize its crest factor \cite{Molisch10}. We transmit three concatenated copies of $x[q]$ from which two copies are used to obtain the oversampled receive signal frequency response while the rest provides a guard period allowing multipath components to decay \cite{Zelenbaba19_2}.

Each sounding signal consists of $Q$ orthogonal frequency division multiplexed (OFDM) subcarriers with a frequency spacing $\Delta f$ and total bandwidth $B=\Delta fQ$. The sounding sequence lasts $t_\mathrm{s}=\frac{3Q}{B}$ and the additional time left $T_\mathrm{s}-t_\mathrm{s}$ after transmitting and receiving allows for buffer clearing, data recording and automatic gain control.

Each receiving node records raw data and estimates the discrete frequency response $g_{(a,b)}[m,q]$ of link $(a,b)$, where $a\neq b$ and $a\in\{ 1,...,L\}$ and $b\in\{ 1,...,L\}$ are the Tx and Rx node indices, respectively. Discrete time and frequency (subcarrier) indices are denoted by $m\in\{ 0,...,T-1\}$ and $q\in\{ 0,...,Q-1\}$, respectively, with $T$ being the total number of recorded time samples and $Q$ the total number of recorded frequency samples. The sampled impulse response is obtained from the sampled frequency response by using the fast Fourier transform.

A calibration measurement between antenna ports of each node pair is performed with the same transmit power as the measurements to obtain the RF chain transfer functions $g_{\mathrm{c}\,(a,b)}[q]$. They are used in post-processing to extract the frequency response as 
\be
g_{(a,b)}[m,q]=y_{b}[m,q]/(x[q]g_{\mathrm{c}\,(a,b)}[q])\,,
\ee
where $y_{b}[m,q]$ is the received signal of node $b$. This ensures wireless channel reciprocity ($g_{(a,b)}[m,q]=g^*_{(b,a)}[m,q]$) and omits the effects of radio frequency (RF) chain components and the Tx signal power.

We compute the time-variant first and second-order statistics of the wireless channel from the recorded frequency responses as described in Appendix~\ref{app:characterization}.

\section{Measurement Campaign}
\label{sec:meas_campaign}
In this section we first present the setup and the measurement parameters used in the measurement campaign. We then present two measurement scenarios and the obtained measurement results, followed by an analysis of observed multipath components.

\subsection{Measurement Setup}

A measurement parameter summary is given in Table~\ref{ChannelSounder_tab}. 

The vehicles in both scenarios are driving at approximately constant speed of maximum $\approx11.5\,\mathrm{m/s}$, making the maximum relative speed $v_\mathrm{max}\approx23\,\mathrm{m/s}$. Therefore, we choose a snapshot duration of $T_\mathrm{sys}=500\,\mathrm{\mu s}$, satisfying the condition $T_\mathrm{sys}<\frac{c_0}{2f_\mathrm{c}v_\mathrm{max}}=1.1\mathrm{ms}$, where $c_0$ is the speed of light and $f_\mathrm{c}$ the carrier frequency, to be able to resolve the largest expected Doppler components. We use $M=240$ samples in time to calculate the channel statistics for each stationarity region, which corresponds to up to $\approx27\,\lambda$ in the spatial domain, or $T_\mathrm{stat}=120\,\mathrm{ms}$ in time. 

We analyze $15\,\mathrm{s}$ of measurement time in each scenario, which corresponds to $T=30000$ samples of measurement data in time. 

\begin{table}[hbtp]
	\begin{center}
		\caption{Channel sounding parameters}
		\begin{tabular}{l l l}
			\toprule
			Parameter    &   Description & Value  \\
			\midrule
			$L$ & number of nodes &	 3	\\
		    $f_\mathrm{c}$    	&carrier frequency     	        & $5.9\,\mathrm{GHz}$         \\
		    $\lambda$       	&wavelength   	                & $5.08\,\mathrm{cm}$        \\
		    
		    $Q$    	            &number of subcarriers     	    & $601$         \\
		    $\Delta f$    	    &subcarrier spacing     	    & $250\,\mathrm{kHz}$         \\
			$B$                 &measurement bandwidth       	& $150.25\,\mathrm{MHz}$       \\
			$f_\mathrm{s}$      &receiver sampling frequency  	 & $160\,\mathrm{MS/s}$       \\
		    $T_\mathrm{sys}$	&snapshot interval 		         & $500\,\mathrm{\mu s}$      \\
		    $T_\mathrm{s}$	    &sounding interval 		         & $250\,\mathrm{\mu s}$      \\
		    $t_\mathrm{s}$	    &sounding sequence length 		 & $12\,\mathrm{\mu s}$      \\
			$v_\mathrm{max}$    &maximum relative velocity of nodes   	     & $\approx$11.5 m/s      \\
			$T_\mathrm{stat}$   &stationarity region duration  	 & $120\,\mathrm{ms}$     \\
			$\Delta\tau$   &delay resolution  	             & $6.67\,\mathrm{ns}$     \\
			$\Delta\nu$    &Doppler resolution  	         & $8.33\,\mathrm{Hz}$     \\
			\bottomrule
		\end{tabular}
		\label{ChannelSounder_tab}
	\end{center}
\end{table}

A summary of hardware setup parameters is given in Table~\ref{HW_setup_tab}. The transmit signal power is fixed to $-10\,\mathrm{dBm}$ such that the output power stays below the $1\,\mathrm{dB}$ compression point of the external amplifier.

\begin{table}[hbtp]
	\begin{center}
		\caption{Hardware setup parameters}
		\begin{tabular}{l l}
			\toprule
			Hardware parameter 	& Value  \\
			\midrule 
			SDR model &	 USRP 2954R	\\
			integrated FPGA model & Xilinx Kintex Series 7 \\
			transmit signal power &	 $-10\,\mathrm{dBm}$	\\
			external amplifier gain &	$27\,\mathrm{dB}$	\\
			maximum automatic gain control gain &	$31.5\,\mathrm{dB}$	\\
			antenna type &	  omni-directional dipole	\\
			antenna polarization &	  vertical 	\\
			antenna gain &	 $4\,\mathrm{dBi}$	\\
			\bottomrule
		\end{tabular}
		\label{HW_setup_tab}
	\end{center}
\end{table}

The vehicle models and antenna heights are given in Table~\ref{Vehicle_tab}. The antennas are mounted on the right front corner of each vehicle roof. Every node is equipped with a dash-cam and a real-time kinematic GPS receiver. The used double-decker bus is $14\,\mathrm{m}$ long, $4\,\mathrm{m}$ high, and $2.55\,\mathrm{m}$ wide.

\begin{table}[hbtp]
	\begin{center}
		\caption{Measurement vehicles}
		\begin{tabular}{l l l}
			\toprule
			Vehicle & Vehicle model	& Antenna height \\
			\midrule 
			$1$ & Toyota Prius  & $1.55\,\mathrm{m}$		\\
			$2$ & Volkswagen Transporter Van & $2.02\,\mathrm{m}$	\\
			$3$  & Toyota Prius	& $1.55\,\mathrm{m}$		\\
			Bus  & Neoplan Skyliner double-decker bus   & No antenna 	\\
			\bottomrule
		\end{tabular}
		\label{Vehicle_tab}
	\end{center}
\end{table}

\subsection{Measurement Scenarios}

The measurement campaign is performed in an urban environment and it includes an overtaking scenario and an intersection scenario. Both scenarios take place in Vienna in Paukerwerkstrasse, which is a two-lane two-way street with office buildings on both sides of the road, a factory at one end, a construction site with metallic fencing further down the road, and trees on the roadside, as shown in the aerial photos in Fig.~\ref{fig:scenario1} and Fig.~\ref{fig:scenario2}.

\subsubsection{Overtaking Scenario}
A diagram of the overtaking scenario is shown in Fig.~\ref{fig:scenario1}. This scenario aims to recreate a situation where car $3$ wants to overtake a large vehicle (in this case a double-decker bus) that is obstructing the LOS to car $1$ which is coming from the opposite direction. Node $2$ is mounted to a van driving in front of the bus and has a LOS connection to car $1$. 

 Another car, driving behind car $1$, is present in the scenario. The distance from the antenna on car $2$ and the bus is kept at $\approx30\,\mathrm{m}$ and the distance between the bus and car $3$ is kept at $\approx10\,\mathrm{m}$, making the distance between antennas of nodes $2$ and $3$ approximately $54\,\mathrm{m}$.

\Figure[!ht]()[width=0.99\columnwidth, height=0.8in]{scenario_1_diagram.png}
{Overtaking scenario diagram with arrows showing the moving direction of vehicles (map source: Google Maps). \label{fig:scenario1}}

\subsubsection{Intersection Scenario}
This scenario recreates a situation where a stable link between two vehicles at opposite sides of an urban intersection is obstructed by a passing large vehicle. As depicted in Fig.~\ref{fig:scenario2} cars $1$ and $3$ are waiting on opposite sides of the road at $30\,\mathrm{m}$ distance, behind building corners, while the bus passes between the two cars, therefore disrupting link $(1,3)$. The bus is followed by car $2$ which is in obstructed line-of-sight (OLOS) with cars $1$ and $3$ while the bus is passing the intersection. 

\Figure[!ht]()[width=0.99\columnwidth]{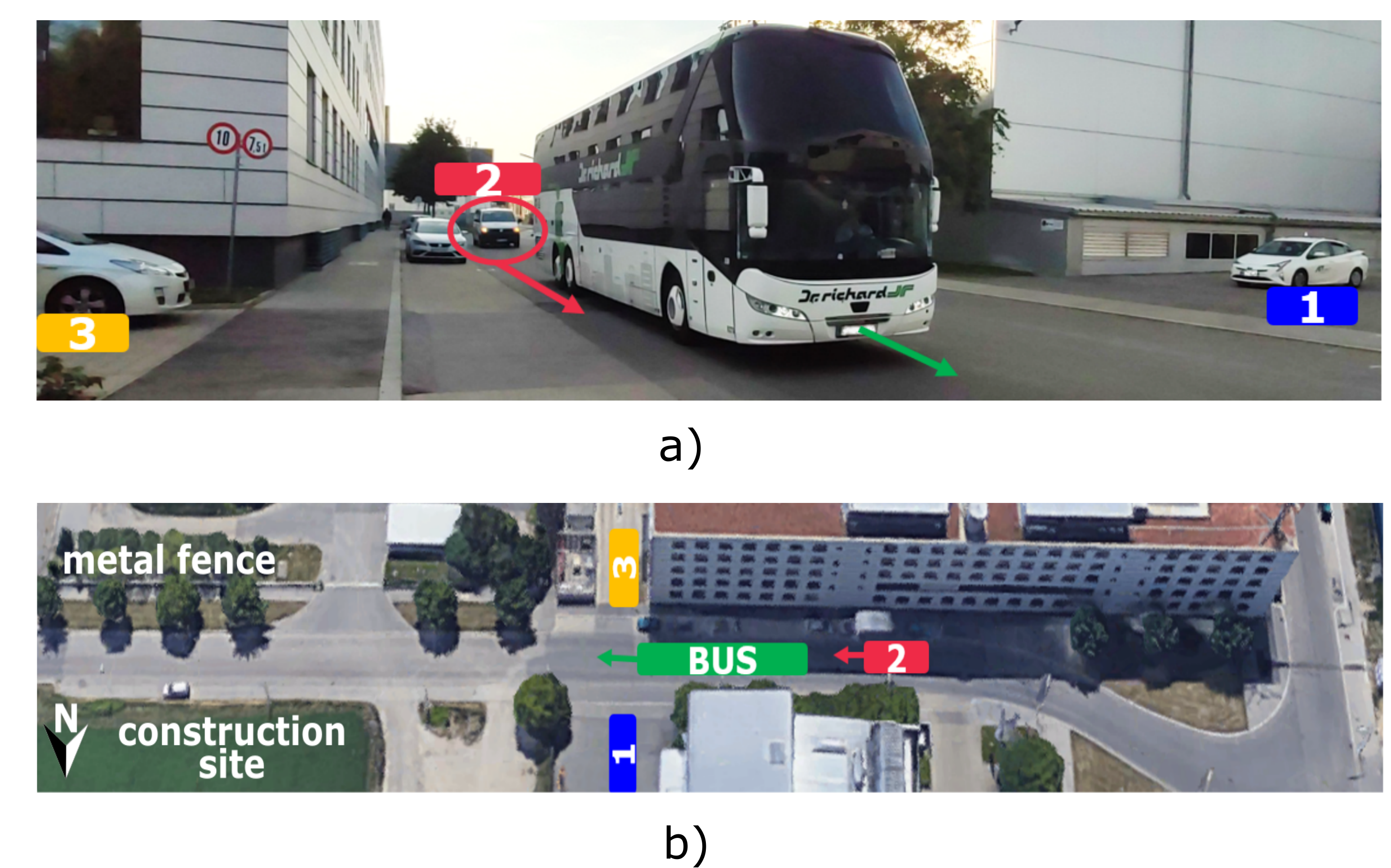}
{Intersection scenario a) photo and b) diagram (map source: Google Maps). \label{fig:scenario2}}

\subsection{Measurement Results}
\subsubsection{Overtaking Scenario}
The time-variant first and second-order statistics of all three wireless channels in the overtaking scenario are shown in Fig.~\ref{fig:scenario1_results} with distinct multipath components pointed out. Links $(1,2)$ and $(1,3)$ show the characteristics of a typical vehicle-to-vehicle communication scenario where vehicles drive in opposite direction and link $(2,3)$ resembles a scenario where both vehicles drive in the same direction \cite{Paier08}. 

After passing the curve at time $t_1=2.3\,\mathrm{s}$, car $3$ becomes shadowed by the bus and the power of the LOS component of link $(1,3)$ drops by $\approx14\,\mathrm{dB}$. Car $1$ passes car $2$ at $t_2=7.9\,\mathrm{s}$ and regains LOS with car $3$. After $t_3=10.1\,\mathrm{s}$ link $(1,2)$ undergoes OLOS conditions. Meanwhile, link $(2,3)$ keeps a steady signal-to-noise ratio of $\approx30\,\mathrm{dB}$ despite the large vehicle obstruction. 

\Figure[t!]()[width=1\linewidth]{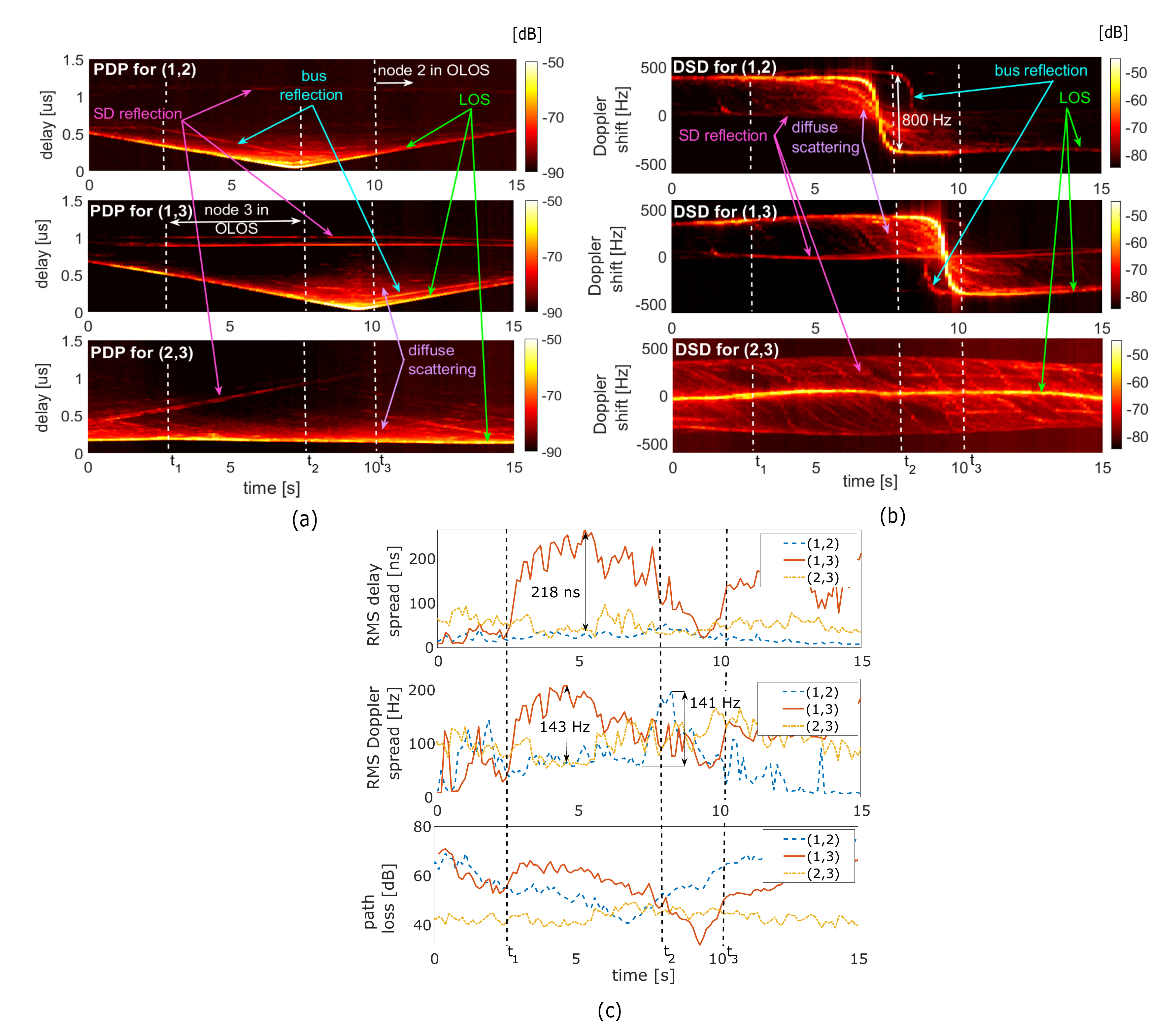}
{   Overtaking scenario:
    (a) Time-variant power delay profiles (PDPs), 
	(b) time-variant Doppler spectral densities (DSDs), and 
	(c) time-variant second-order statistics, of all three measured channels in the overtaking scenario.
	Link $(1,3)$ is in obstructed line-of-sight (OLOS) from $t_1$ until $t_2$, and link $(1,2)$ becomes obstructed after $t_3$.
	\label{fig:scenario1_results}}

\subsubsection{Intersection Scenario}
In the intersection scenario, links $(1,2)$ and $(2,3)$ present typical vehicular communication links of two vehicles passing each-other \cite{Paier08}. Therefore we focus on the analysis of the disrupted link $(1,3)$. The time-variant power delay profile (PDP) and Doppler spectral density (DSD) of link $(1,3)$ are shown in Fig.~\ref{fig:scenario2_pdpdsd}. The bus obstructs link $(1,3)$ between $t_4=6.1\,\mathrm{s}$ and $t_5=7.9\,\mathrm{s}$ and the van with node $2$ then obstructs the link between $t_6=9.5\,\mathrm{s}$ and $t_7=10.3\,\mathrm{s}$.

The static discrete (SD) scatterer reflection component pointed out in Fig.~\ref{fig:scenario1_results} comes from the metallic structures on the factory marked at the right end of Fig.~\ref{fig:scenario1}. Comparing to link $(1,3)$, this component is weaker by roughly $17\,\mathrm{dB}$ in link $(1,2)$ as it has to go through the bus to reach car $2$. After accounting for the path loss of the additional distance between nodes $2$ and $3$, we can see that the bus attenuates the SD reflection by $\approx13\,\mathrm{dB}$. 

Other SD reflections coming from metallic structures such as traffic signs are also present in the results. The diffuse scattering reflections coming from buildings, parked cars and vegetation are present in all three links, as shown in Fig.~\ref{fig:scenario1_results}.

\subsubsection{Large Vehicle Impact}
As car $1$ approaches car $2$ in the overtaking scenario, the signal of link $(1,2)$ is reflected from the front side of the bus and then in link $(1,3)$ from the back of the bus, $0.48\,\mathrm{s}$ later, as seen in  Fig.~\ref{fig:scenario1_results}. The back reflection is noticeably stronger than the front reflection since the back side of the bus has a larger metallic surface. This hypothesis is confirmed in the intersection scenario where the back of the bus causes a stronger reflection than the front side. 

The obstructing large vehicle has a strong influence on the RMS delay spreads. In the overtaking scenario the difference in the RMS delay spread of the obstructed and LOS link is more than $200\,\mathrm{ns}$, as shown in Fig.~\ref{fig:scenario1_results} c). With an increase in distance between the vehicles, the large vehicle reflections would proportionally increase the RMS delay spreads.

Between $t_2$ and $t_3$, the Doppler shift of the reflection coming from the front of the bus and the LOS component of link $(1,2)$ is more than $800\,\mathrm{Hz}$. This causes the RMS Doppler spread to increase by more than $140\,\mathrm{Hz}$, as seen in Fig.~\ref{fig:scenario1_results}. Furthermore, the difference between the RMS Doppler spreads of links $(1,3)$ and $(1,2)$ between $t_1$ and $t_2$ is also more than $140\,\mathrm{Hz}$. This difference would further increase at higher velocities and the impact of the bus reflections on the RMS Doppler spreads would be even more severe.

During the OLOS interval in the intersection scenario, the OLOS signal fades more as the bus moves forward, as seen in Fig.~\ref{fig:scenario2_pdpdsd}, indicating the different signal penetration of different parts of the bus.

As seen in Fig.~\ref{fig:scenario2_rms} the RMS delay spread increases by $30\,\mathrm{ns}$ and the path loss by $8.6\,\mathrm{dB}$ right after $t_4$ when the front of the bus, which is largely covered in glass, is obstructing the link. The RMS delay spread further increases by up to $69\,\mathrm{ns}$ when the back of the bus, which is mostly metallic, is obstructing the link and the path loss is further increased by up to $20\,\mathrm{dB}$. 

The RMS Doppler spread in the intersection scenario increases by $43\,\mathrm{Hz}$ while the bus is obstructing the link, and increases by up to $51\,\mathrm{Hz}$ between $t_5$ and $t_6$ owing to the combined impact of reflections from both the bus and car $2$. 
The van carrying node $2$ causes an increase of up to $35\,\mathrm{ns}$ in the RMS delay spread, $18.7\,\mathrm{Hz}$ RMS Doppler spread and up to $7.5\,\mathrm{dB}$ path loss between $t_6$ and $t_7$, significantly less than the bus.

\Figure[!ht]()[width=0.99\columnwidth]{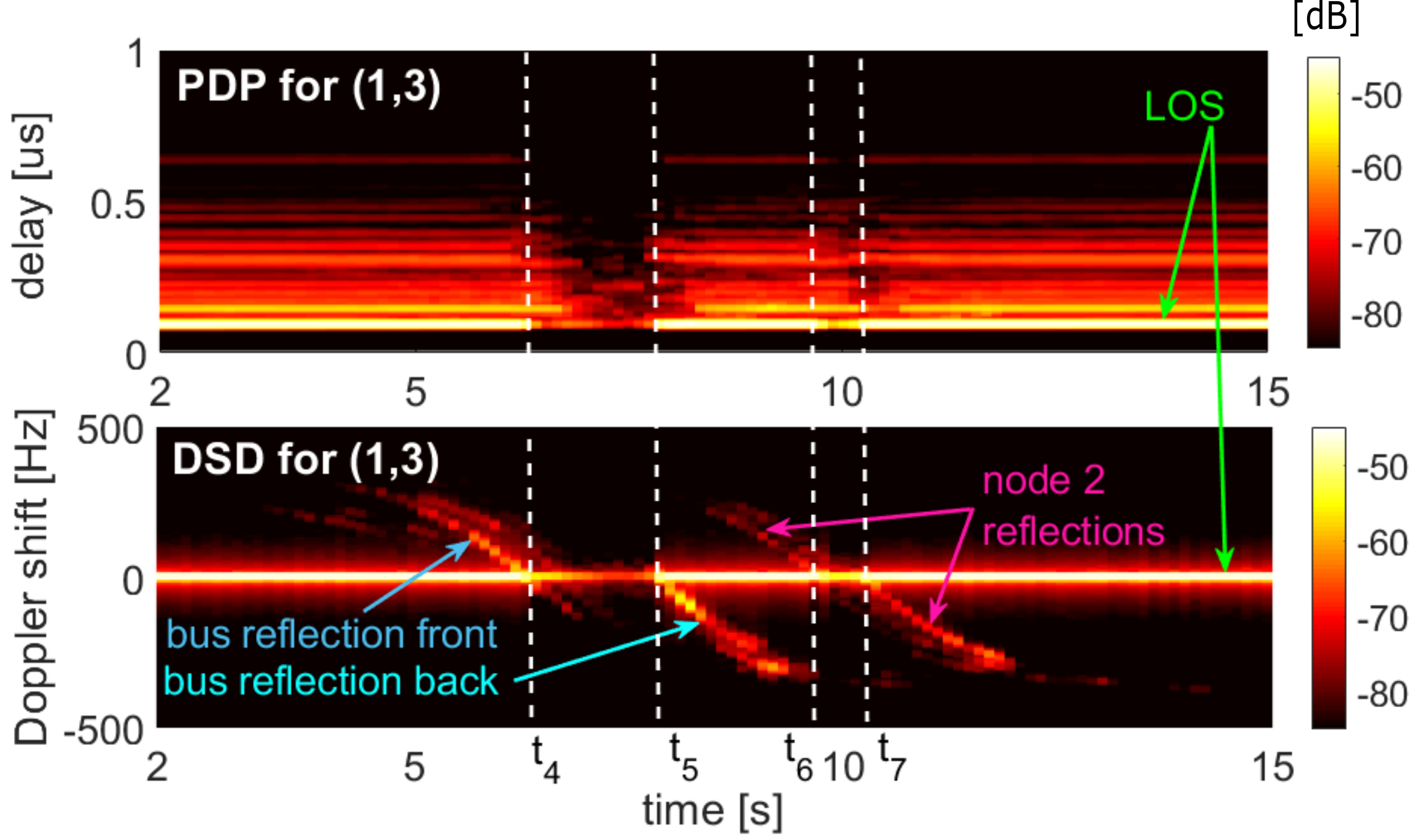}
{Intersection scenario: Time-variant PDP and DSD of link $(1,3)$. The bus obstructs the link between time instants $t_4$ and $t_5$, and car $2$ obstructs the link between $t_6$ and $t_7$ \label{fig:scenario2_pdpdsd}}

\Figure[!ht]()[width=0.99\columnwidth]{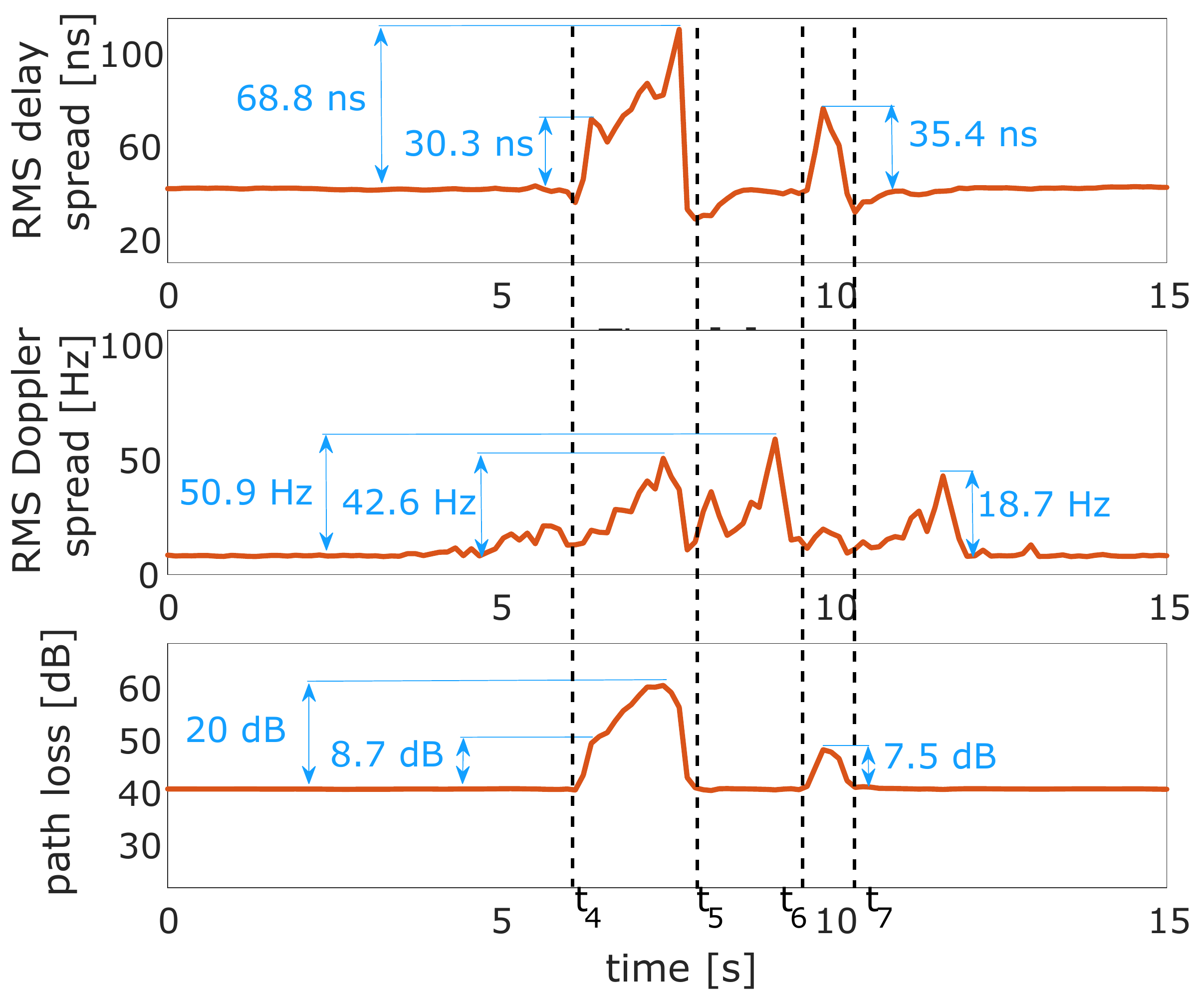}
{Intersection scenario: Time-variant RMS delay spread, RMS Doppler spread and path loss of link $(1,3)$. \label{fig:scenario2_rms}}

\section{Geometry-Based Channel Model}
\label{sec:gscm}
In order to model the time-variant second-order statistics of the measured wireless channels, we recreate the measurement environment geometry and the path coefficients by using a GSCM. As mentioned in Section~\ref{sec:related_work}, we base our model on the model presented in \cite{Karedal09}. 

Using a path-based GSCM together with publicly available data from OSM yields a very high level of flexibility when it comes to modelling the scenario geometry. This modelling approach can be used to apply the model parameters to different scenarios and easily scale to system-level \cite{Dakic_21}.

In this section we provide a short description of the methodology used for developing our OSM-GSCM, while a more detailed description can be found in \cite{Rain2004:Optimized}. The insights gathered in two measurement scenarios are used to model the impact of large obstructing vehicles, which is the main novelty of our channel model.

\subsection{Geometry Modelling}
Most of the scenario geometry data such as the position of buildings, traffic signs, roads, and vegetation is imported from OSM, thus significantly reducing modeling efforts. Node trajectories are imported from the recorded GPS data, counting in the relative position of the GPS antennas to the transceiver antennas. We model the bus and the car driving behind car $1$ as mobile discrete scatterers and input their trajectories manually. All the trajectories are sampled once per second and the coordinates of the vehicles between the sample points are spline-interpolated. 

Diffuse scatterers are uniformly distributed along building walls according to predefined distributions. A uniformly distributed initial phase for each diffuse scatterer is selected for each simulation run. To speed-up the simulation we use only a limited number of diffuse scatterers selected through the locality-sensitive hashing algorithm \cite{Rain2004:Optimized}. Attenuation of the LOS signal by vegetation is determined according to \cite{Adegoke2016VegetationAA}. Metallic reflecting objects such as traffic signs are modeled as SD scatterers supporting first order reflections, to keep the computational complexity of the model low. A screenshot of the simulation geometry for the overtaking scenario is shown in Fig.~\ref{fig:simulation_geometry}. 

A summary of model parameters for different scatterer types is given in Table~\ref{tab:model_parameters}. We use the measurement results to calibrate this small set of model parameters to optimize the match between measurements and simulation.

\Figure[!ht]()[width=0.99\columnwidth, height=1.5in]{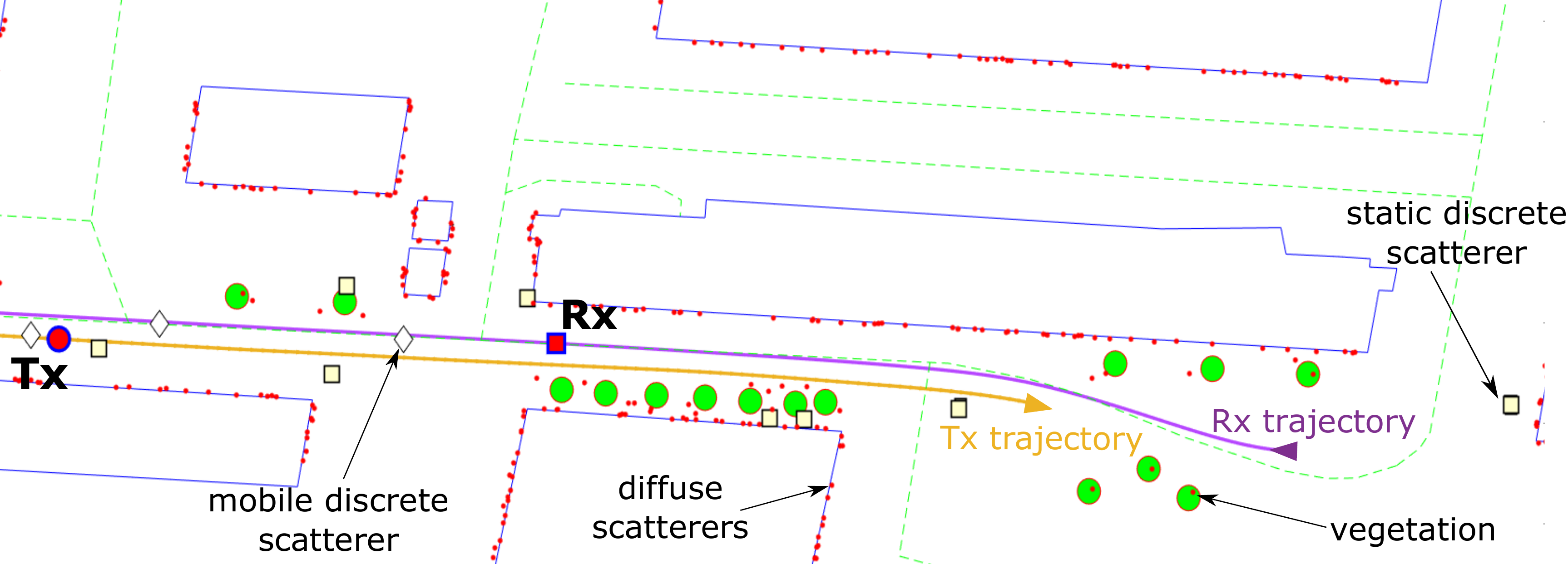}
{Overtaking scenario: Geometry-based stochastic channel model (GSCM) simulation geometry using data imported from OpenStreetMap (OSM) and GPS trajectories from the measurements. \label{fig:simulation_geometry}}

\begin{table}[hbtp]
	\begin{center}
		\caption{Channel model parameters \cite{Karedal09} for the line-of-sight (LOS) component, static discrete scatterers (SD), mobile discrete scatterers (MD) and diffuse scatterers (DI).}
		\begin{tabular}{p{1.2cm} p{2.5cm} l l l l}
			\toprule
		 	Parameter & Description & LOS & SD & MD & DI  \\
			\midrule 
	        $G_0$ [dB]	           &   reference Rx power          &	 $-37$	& $-89$ &	$-97$	& $-39$	\\
	        $n_\mathrm{p}$ 	           &   path loss exponent          &	 $1.9$	& $1.5$ &	$3.6$	& $3.3$	\\
	        $\mu_\sigma$ [m]	   &    mean variance of the stochastic amplitude gain         &	 $1$	& $3.1$ &	$3.13$	& - \\
	        $\mu_c$ [m]	           &     mean coherence distance       &	 $1.2$	& $4.9$ &	$5.4$	& -	\\
	        $d_c^{\mathrm{min}}$ [m]&	  minimum coherence distance      &	 $1.4$	& $1$   &	$1.1$	& -	\\
	        $\chi$ $\mathrm{[m^{-1}]}$	&   scatter point density     &	 -	    & $0.3$ &	$0.01$	& $0.5$	\\
			$w$ $\mathrm{[m]}$	&    maximum scatterer distance    &	 -	    & $0.3$ &	$0.01$	& $0.5$	\\
			
			\bottomrule
		\end{tabular}
		\label{tab:model_parameters}
	\end{center}
\end{table}

\subsection{Large vehicle model}
The authors of \cite{Vlastaras17} have shown the contribution of the double edge-diffracted signal over the roof of a truck. However, large passenger vehicles tend to have a larger percentage of windows and a more heterogeneous structure. This leads to different penetration loss, depending on the part of the large vehicle that is obstructing the LOS, as we have seen in Fig.~\ref{fig:scenario2_rms}. 

The authors of \cite{Shu18} show that the side of a bus passing parallel to the measuring nodes, behaves as a moving reflecting surface. From our analysis of Fig.~\ref{fig:scenario2_pdpdsd}, we see that the front and back of a large moving vehicle show the same behaviour. 

Therefore we focus on including two major factors of large vehicle contributions in our model:
\begin{itemize}
    \item The reflections from the different sides of the large vehicle - when Tx and Rx are on the same side of the large vehicle.
	\item The path loss of an obstructing large vehicle - when Tx and Rx are on opposite sides of the large vehicle.
\end{itemize}  
  
\subsubsection{Side Reflection}
With the aim of keeping our model computationally simple we consider only a single point reflection on the middle axis of the large vehicle, but we change the position of the reflection point $x_\mathrm{MD}$ depending on the position of the nodes relative to the large vehicle. Therefore we consider three separate cases, as depicted in Fig.~\ref{fig:MD_angles}:
\vspace{3mm}
\begin{equation}
x_\mathrm{MD}=
	\begin{cases}
    \frac{l_\mathrm{MD}}{2},& \mid \theta_\mathrm{Tx\,front} \mid \leq \frac{\pi}{2} \; \land 
    \mid \theta_\mathrm{Rx\,front} \mid \leq \frac{\pi}{2}\\
    -\frac{l_\mathrm{MD}}{2},& \mid \theta_\mathrm{Tx\,back} \mid \geq \frac{\pi}{2} \; \land
    \mid \theta_\mathrm{Rx\,back} \mid \geq \frac{\pi}{2} \\
    0,& \mathrm{otherwise}
    \end{cases}
\end{equation}
where $\theta_\mathrm{Tx\,front}$ and $\theta_\mathrm{Rx\,front}$ are angles between the Tx and the Rx and the middle point of the front side of the large vehicle, respectively, relative to its direction. The angles between the middle point of the back side of the large vehicle and the Tx and the RX, relative to the direction of the large vehicle, are denoted by $\theta_\mathrm{Tx\,front}$ and $\theta_\mathrm{Rx\,front}$, respectively.

\Figure[!ht]()[width=0.99\columnwidth]{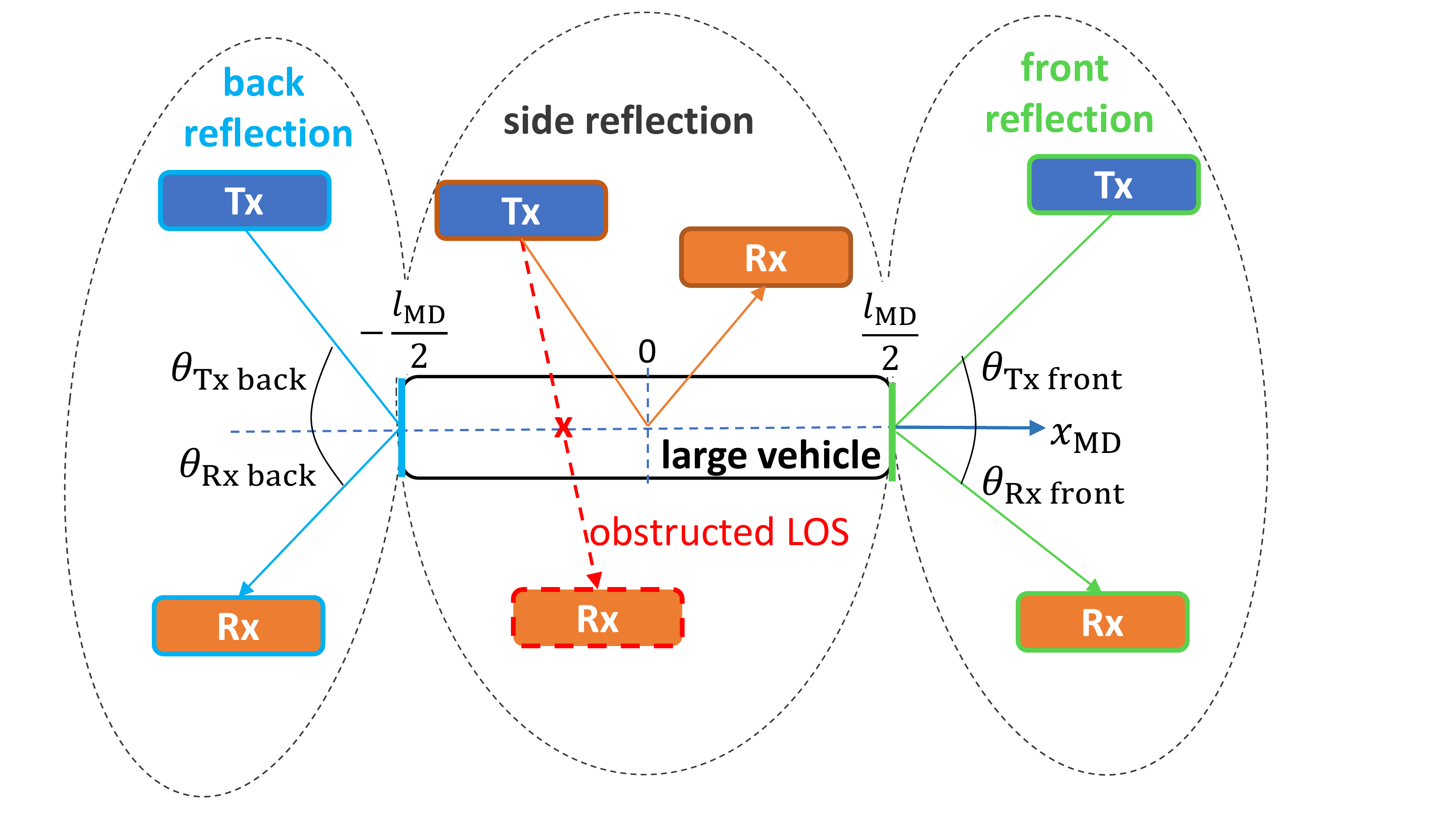}
{The reflection point is selected depending on the position of the nodes relative to the large vehicle. Obstructed line-of-sight (OLOS) is considered in case the nodes are not positioned on the same side of the large vehicle. The figure depicts the Tx in different positions, connected by the solid-line reflection path (or obstructed path marked by the dashed line) to the corresponding position of the Rx. \label{fig:MD_angles}}

In case of our bus, the front and the back side have a different size of the metallic reflective surface and glass. The size of these metallic surfaces is larger than in a typical passenger car, hence they cause proportionally stronger reflections. Due to the two-dimensional nature of the GSCM and its use of point reflections we use the measurement data from the intersection scenario to quantify this difference and accordingly add $12\,\mathrm{dB}$ gain to the back reflection of the bus and an additional $7\,\mathrm{dB}$ gain to the front reflection. For the same reason we add an additional $4\,\mathrm{dB}$ gain to the back reflection of the node $2$ van.

\subsubsection{Path Loss}

Depending on the intersecting point of the LOS path and the horizontal axis $x_\mathrm{MD}$ of the obstructing vehicle, we apply the normalized measured path loss of the LOS component from the intersection scenario between times $t_4$ and $t_5$, to the simulated LOS path. The simulated path loss of the intersection scenario is compared to three measurement runs in Fig.~\ref{fig:bus_pl}.

\Figure[!ht]()[width=0.99\columnwidth]{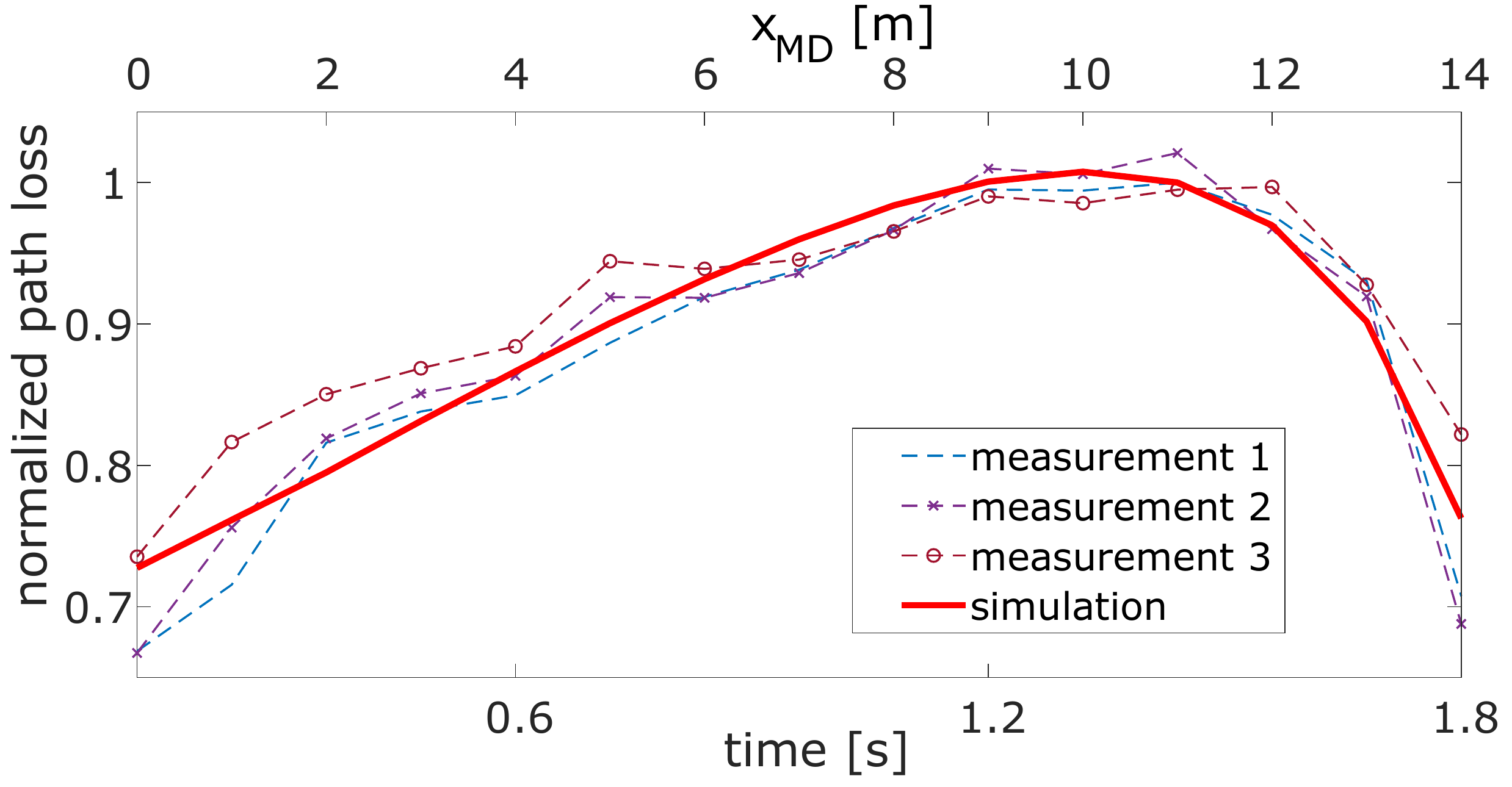}
{Measured and simulated normalized path loss during the double decker bus obstruction in the intersection scenario. The attenuation factor is applied to the LOS component depending on the intersection point of the horizontal axis of the large vehicle $x_\mathrm{MD}$ and the LOS path.\label{fig:bus_pl}}

When the LOS intersects both the front and the back side of the obstructing large vehicle, we add an attenuation factor of $\alpha_\mathrm{bus}[m]=0.008\,d_\mathrm{LOS}[m]$ to the path loss exponent, where $d_\mathrm{LOS}[m]$ is the LOS distance between the Tx and the Rx.

Due to the path-based nature of our model, the approach to modelling large vehicles proposed in this paper can easily be scaled to scenarios with multiple large vehicles by applying the same modelling principle to each large vehicle individually.

\subsection{Simulation results}

\subsubsection{Overtaking Scenario}
The PDP and DSD of the simulated link are shown in Fig.~\ref{fig:sim_PDP_DSD}. The simulated results show a good qualitative match with the measured PDP and DSD of link $(1,3)$ shown in Fig.~\ref{fig:scenario1_results}. 

\Figure[!ht]()[width=0.99\columnwidth]{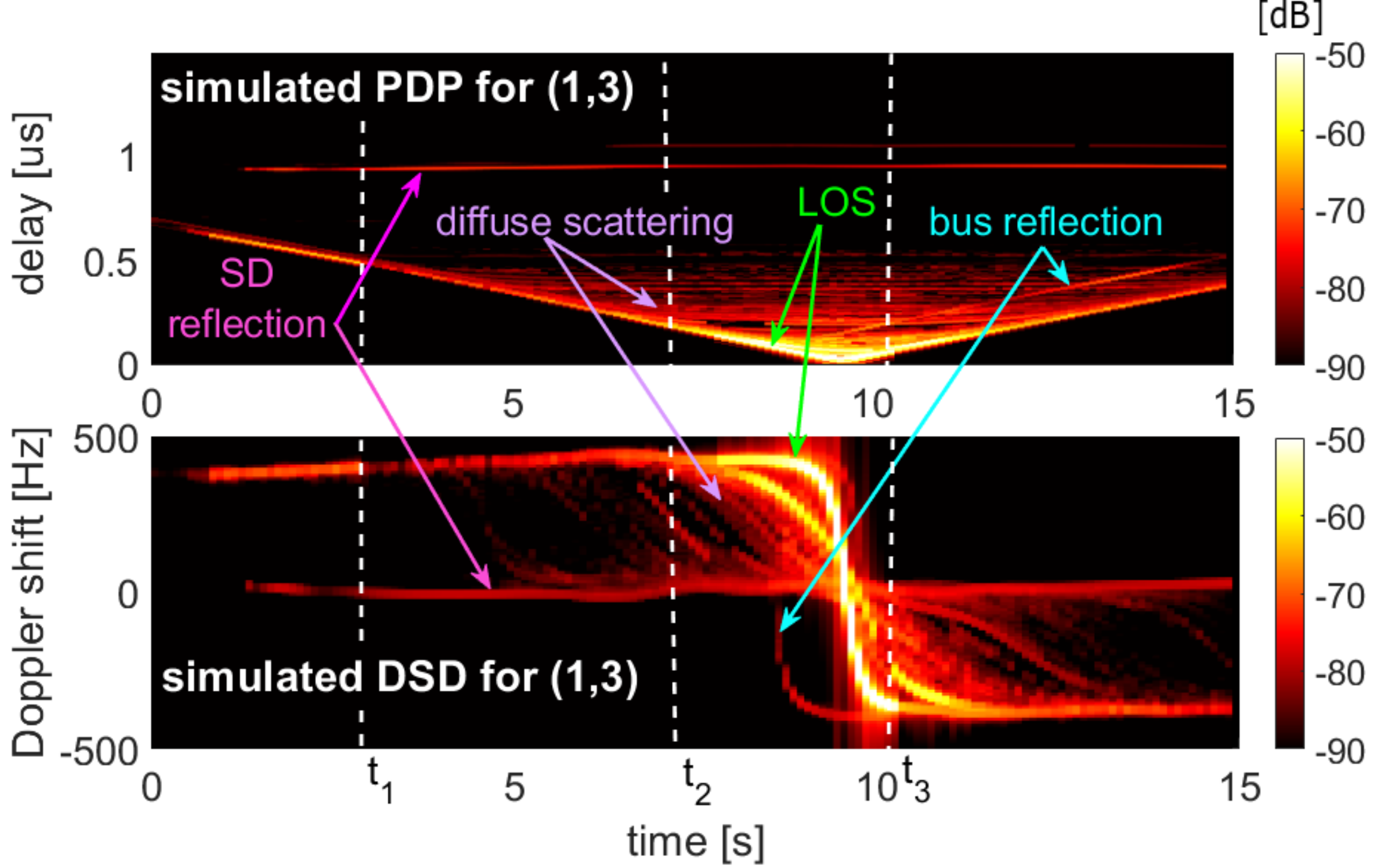}
{Overtaking scenario: PDP and DSD of link $(1,3)$ from the OSM-GSCM simulation. \label{fig:sim_PDP_DSD}}

\Figure[!ht]()[width=0.99\columnwidth, height=3in]{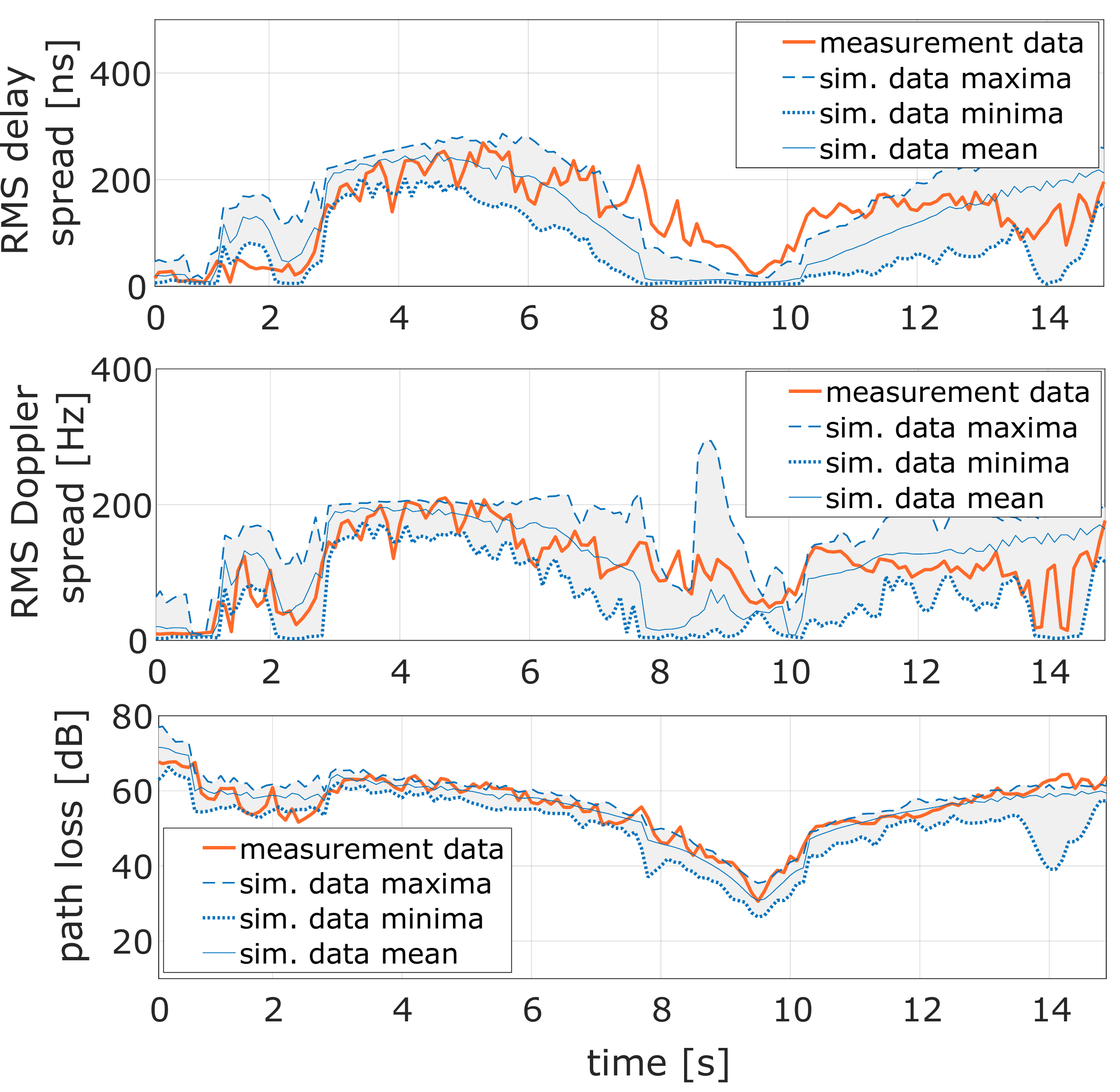}
{Overtaking scenario: Measured and simulated second-order statistics of link $(1,3)$. The area between the minima and the maxima of 100 simulation runs is shaded grey. \label{fig:sim_RMS_PL}}

A comparison of the measured second-order statistics with the minima and maxima of second-order statistics of 100 simulation runs (with randomly initiated diffuse scatterer phases) is shown in Fig.~\ref{fig:sim_RMS_PL}. The simulated second-order statistics also provide an excellent match as measurement results fall mostly between the minima and the maxima of the simulation runs.

\subsubsection{Quantitative Comparison}
To quantify the match between the model and the measurement we calculate the difference between the estimated mean of the second-order statistics obtained by the simulation runs and the ones obtained from the measurement data. The cumulative distribution functions of the calculated offset values are shown in Fig.~\ref{fig:abs_offset_cdf}. The path loss deviates less than $3.6\,\mathrm{dB}$ for $80\%$ of the total simulation duration. The RMS delay spread and RMS Doppler spread offset is less than $78\,\mathrm{ns}$ and $52\,\mathrm{Hz}$, respectively, for $80\%$ of the samples.

\Figure[!ht]()[width=0.99\columnwidth]{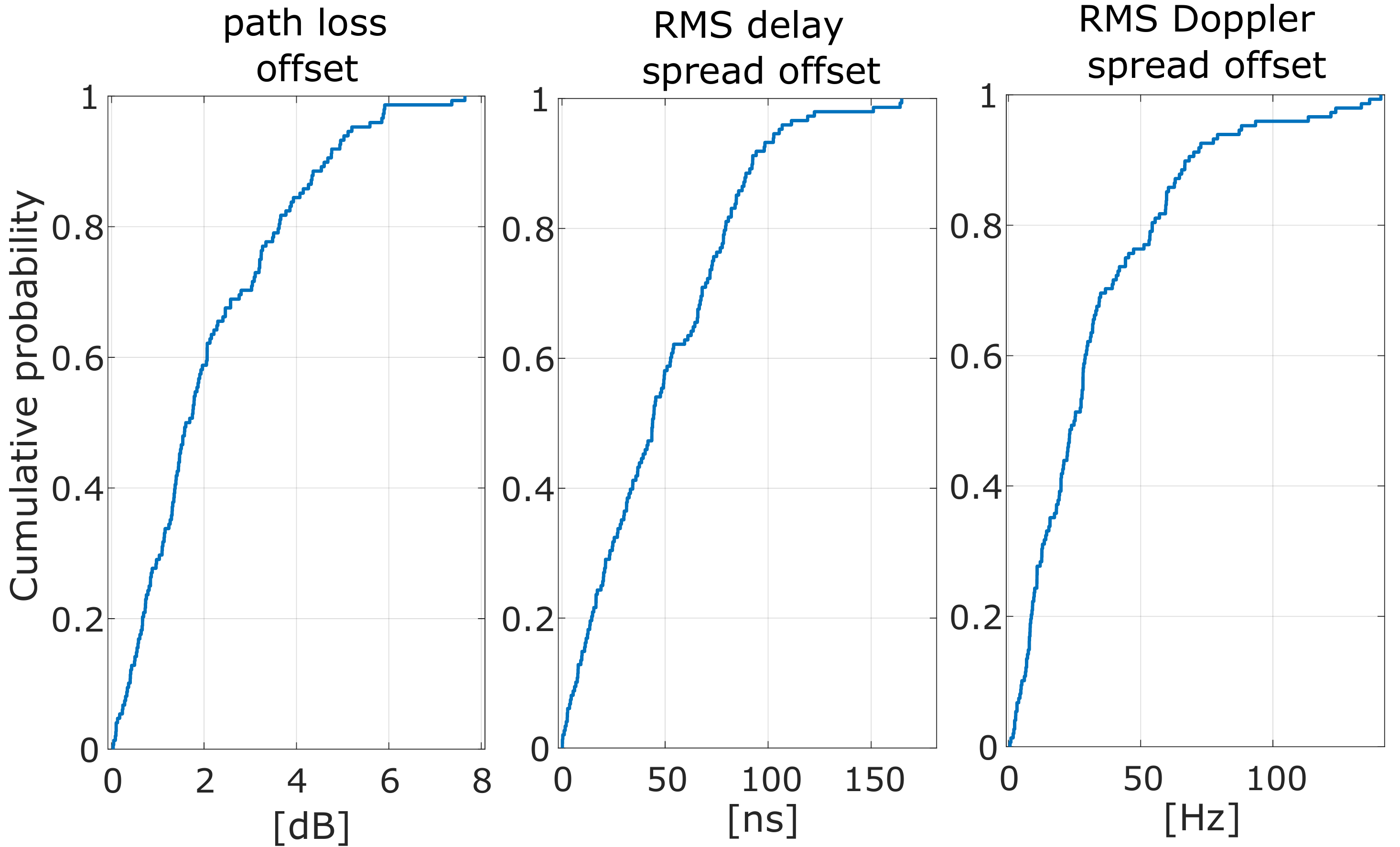}
{Overtaking scenario: Cumulative distribution functions of offset values of the mean second-order statistics obtained from simulations to the measurement data. \label{fig:abs_offset_cdf}}

\subsubsection{Intersection Scenario}
The time-variant PDP and DSD obtained by simulating the intersection scenario are shown in Fig.~\ref{fig:sim_PDP_DSD_scenario2}, displaying a good match with the measurement results from Fig.~\ref{fig:scenario2_pdpdsd}. However, there are additional reflective components that occur in the measurements due to the size and exact shape of the reflective surfaces of the obstructing vehicles.

\Figure[!ht]()[width=0.99\columnwidth]{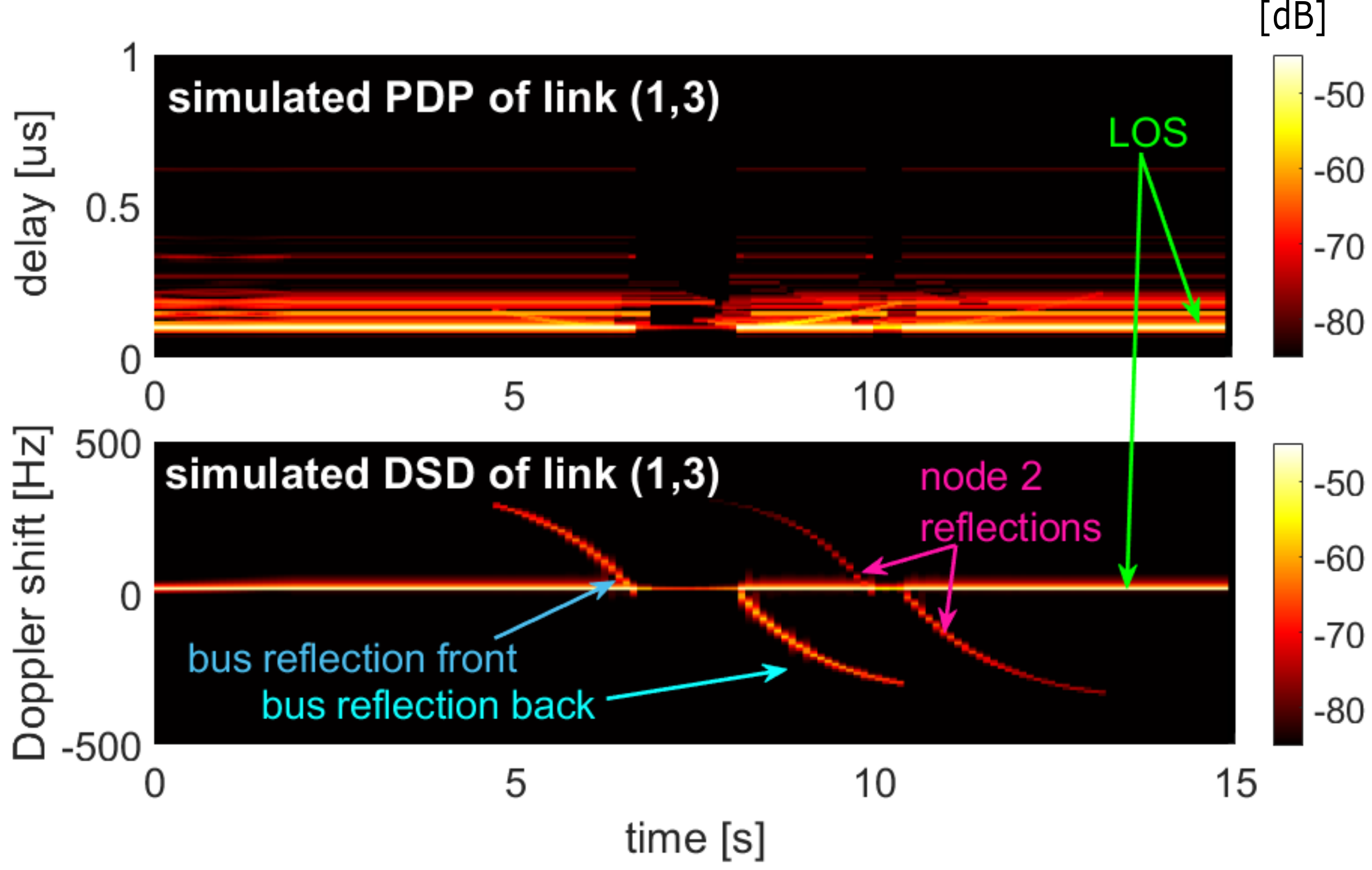}
{Intersection scenario: PDP and DSD of link $(1,3)$ from the OSM-GSCM simulation. \label{fig:sim_PDP_DSD_scenario2}}

In both scenarios we show an excellent match between the measurement results and simulation results but minor offsets still occur due to the absence of three dimensional components and higher order reflections from the model, the modeling of large reflective surfaces as point scatterers, and the exclusion of minor objects such as parked vehicles, metallic drainpipes or fences. Further improvements could therefore be achieved, albeit at the cost of drastically increasing the computational complexity of the model.

\section{Packet Error Rate Evaluation}
\label{sec:per_emulation}
As an additional step to verify the channel model we use a hardware-in-the-loop (HiL) methodology that allows us to obtain high resolution time-variant PERs from measured frequency responses \cite{Zelenbaba20_2}. We use the measured and the simulated frequency responses as input to the AIT real-time wireless channel emulator \cite{Hofer19} and then compare the PER of wireless links emulated between two off-the-shelf IEEE 802.11p \cite{80211p} modems. The diagram of the HiL setup is depicted in Fig.~\ref{fig:fer_meas_setup}.

\subsection{Link-level Emulation Setup}
For our link-level emulation, we use off-the-shelf Cohda Wireless MK5 modems \cite{Cohda} as transceivers. The emulator \cite{Hofer19} is based on the basis expansion model approach and consists of a propagation module, implemented on a host PC, and a convolution module implemented on the FPGA of a USRP SDR, as shown in Fig.~\ref{fig:fer_meas_setup}.

\Figure[!ht]()[width=0.99\columnwidth]{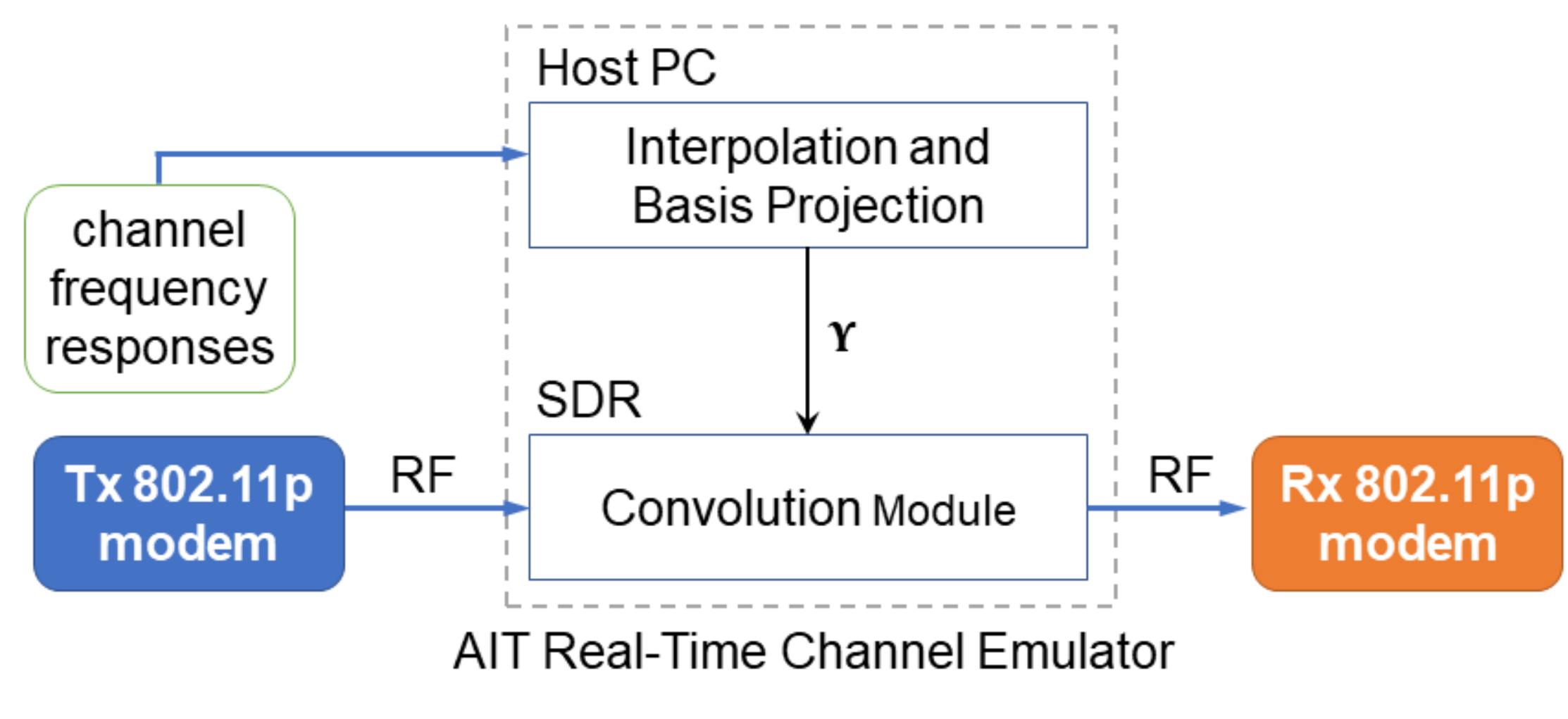}
{Hardware-in-the-loop (HiL) packet error rate (PER) measurement setup using the AIT real-time channel emulator. \label{fig:fer_meas_setup}}

The emulated wireless communication channels have a bandwidth of $10\,\mathrm{MHz}$ at a center frequency of $5.9\,\mathrm{GHz}$ (channel 180). To adjust for different sampling in time and frequency between input frequency responses and emulation, we use interpolation based on discrete prolate spheroidal sequences \cite{Slepian78}. A mathematical description of the used interpolation method is given in Appendix~\ref{app:interpolation}. The interpolated frequency responses are then projected on the used basis sequences (also discrete prolate spheroidal sequences) and the obtained coefficients are streamed to the convolution module. The convolution module reconstructs the frequency responses, convolves them with the signal coming from the Tx modem and forwards it to the Rx modem.

The presented approach allows for a significantly reduced streaming bandwidth compared to streaming the uncompressed frequency responses. It also allows easy and repeatable link-level testing for different modems and with different communication parameters.

In our setup, we use a transmit power of $0\,\mathrm{dBm}$ and a QPSK symbol alphabet with a convolutional coding rate of $1/2$. The packet size is set to the typical IEEE 802.11p size of $100$ bytes, and we transmit at a rate of $1000\,\mathrm{packets/s}$, which gives us a throughput of $800\,\mathrm{kbit/s}$. To reach very low PERs of down to $10^{-4}$ we consider the packet error probability of $100$ combined emulation runs.

\subsection{PER analysis}

In Fig.~\ref{fig:fer_analysis}, the PER obtained by emulating empirically measured frequency responses of channel $(1,3)$ in the overtaking scenario, denoted by $\gamma[k]$, is compared to the PERs obtained by emulating frequency responses from 100 OSM-GSCM simulation runs with randomly initialized diffuse scatterer phases. The estimated mean PER of emulated simulations is denoted by $\bar{\gamma}'[k]$ and the minima and the maxima are denoted by $\gamma'_\mathrm{min}[k]$ and $\gamma'_\mathrm{max}[k]$, respectively.

\Figure[!ht]()[width=0.99\columnwidth]{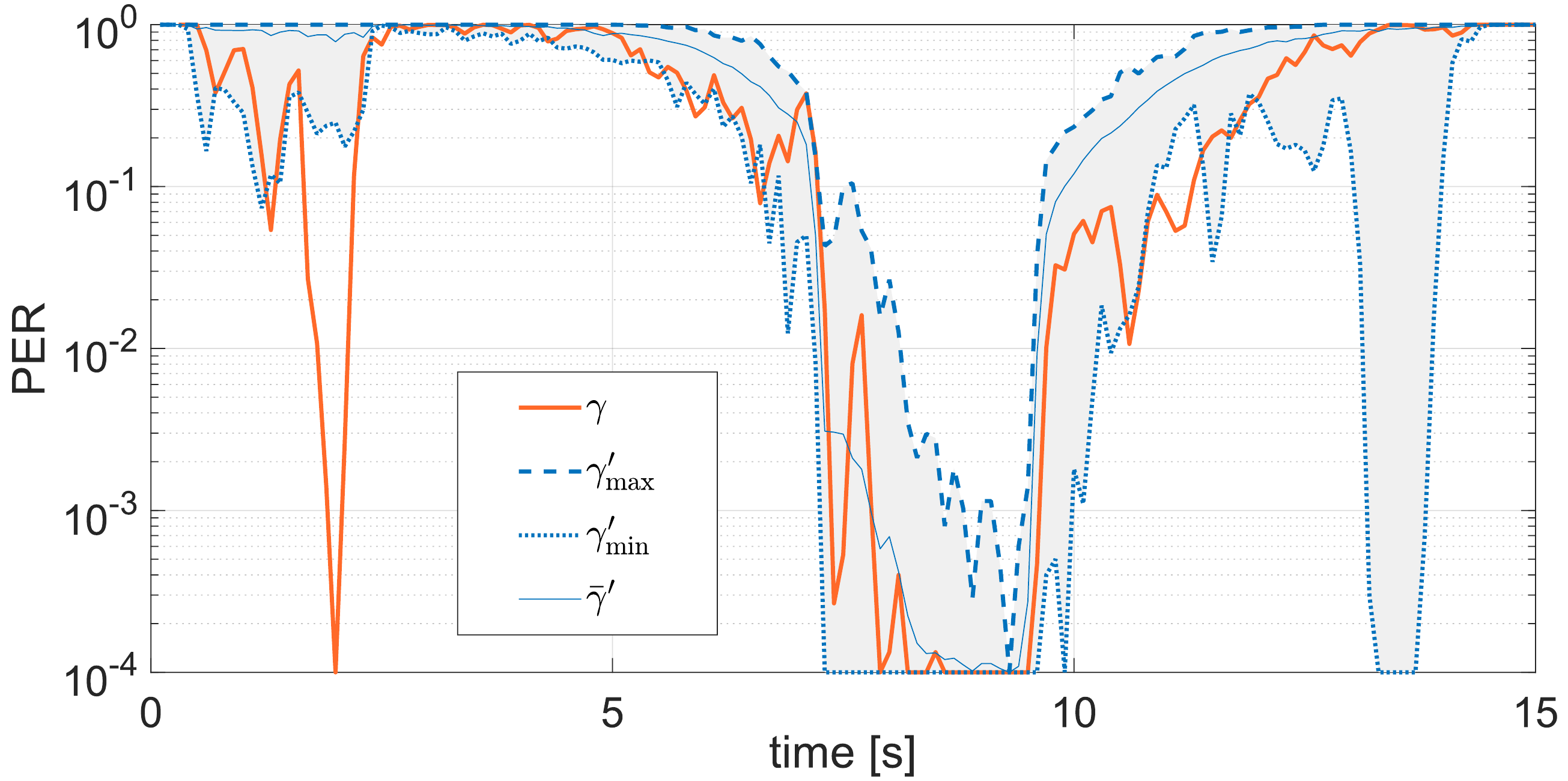}
{Overtaking scenario: Time-variant PER $\gamma$ obtained by emulating the measured channel $(1,3)$, and the estimated mean PER ($\bar{\gamma}'$), maxima ($\gamma'_{\mathrm{max}}$), and minima ($\gamma'_{\mathrm{min}}$) of PERs obtained from emulating 100 simulation runs. The used modulation is QPSK with a convolutional coding rate of $1/2$. \label{fig:fer_analysis}}

The value of $\gamma[k]$ falls between $\gamma'_\mathrm{min}[k]$ and $\gamma'_\mathrm{max}[k]$ for $86\%$ of the time, showing that our OSM-GSCM consistently provides a good match with the measurements, also at the link-level. Furthermore, by calculating the offset ratio $\gamma[k]/\bar{\gamma}'[k]$ we see that the ratio is kept between $0.09$ and $1$ for $90\%$ of the values, as shown in Fig.~\ref{fig:offset_ratio_cdf}.

\Figure[!ht]()[width=0.99\columnwidth]{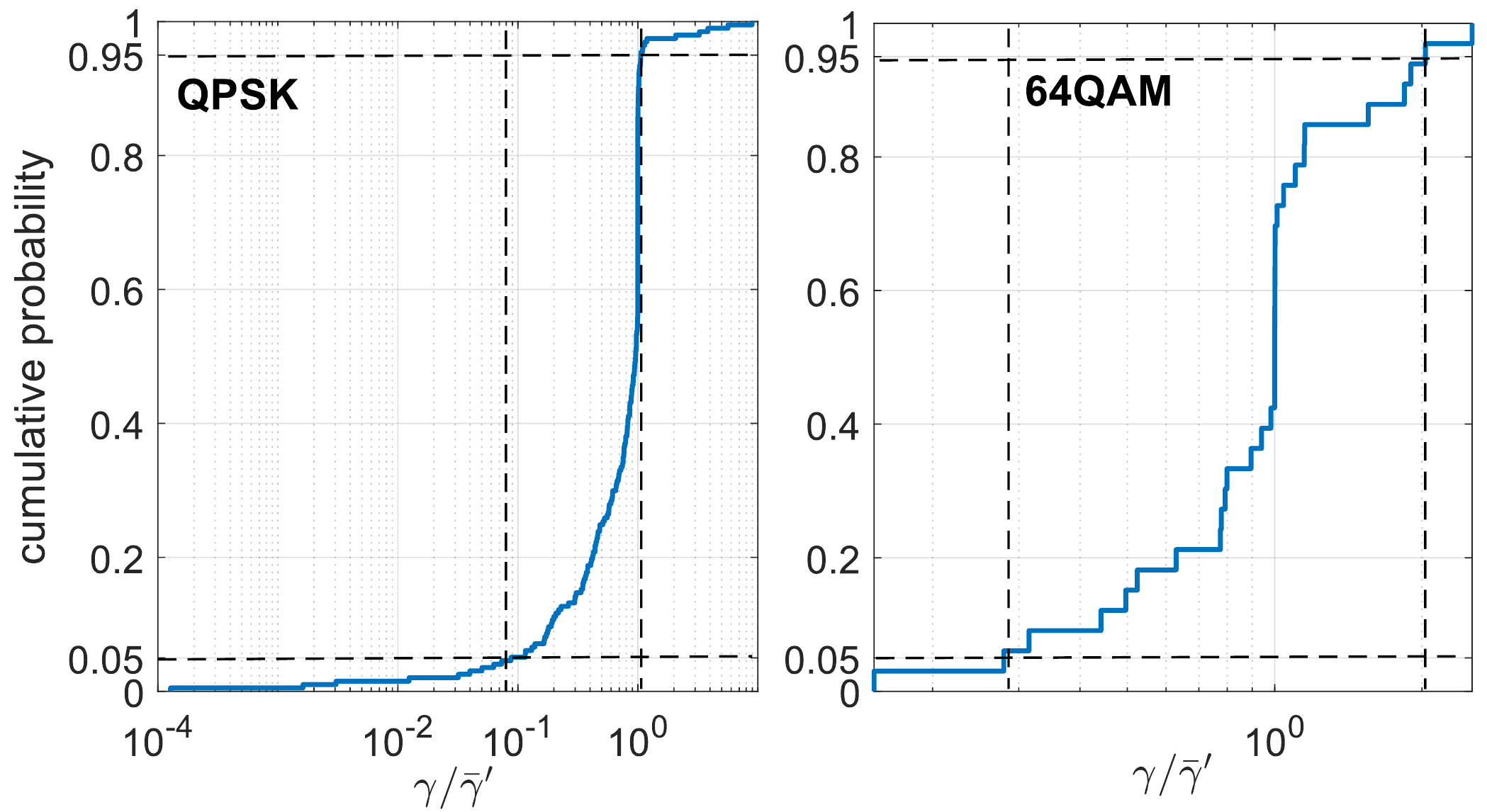}
{Cumulative distribution functions of ratios between PER $\gamma[k]$, obtained from emulating measured frequency responses, and the estimated mean PER $\bar{\gamma}'$, obtained from emulating simulated frequency responses, for both used modulation schemes. \label{fig:offset_ratio_cdf}}

We then run the emulations with the same parameters but with a 64QAM symbol alphabet and a convolutional coding rate of $3/4$. The result is presented in Fig.~\ref{fig:fer_analysis_64qam} where the PER $\gamma[k]$ obtained from measurement data falls between the $\gamma'_{\mathrm{min}}[k]$ and the $\gamma'_{\mathrm{max}}[k]$ in $98\%$ of the emulated time, showing an excellent match of our model for more complex modulation schemes. As shown in Fig.~\ref{fig:offset_ratio_cdf}, the offset ratio in case of 64QAM is found to be between $0.27$ and $2$ for $90\%$ of the values of the analyzed interval.

\Figure[!ht]()[width=0.99\columnwidth]{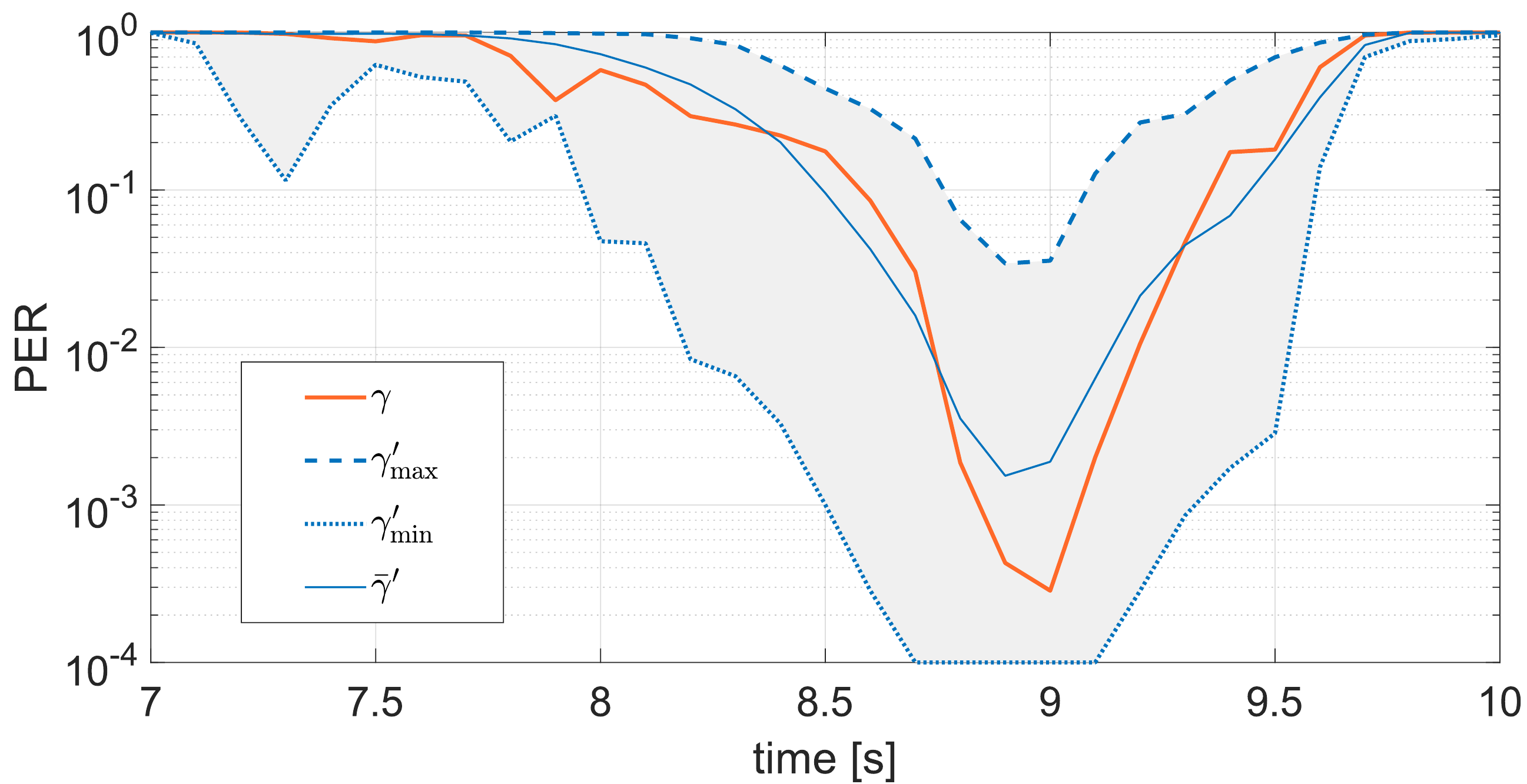}
{Overtaking scenario: Time-variant PER $\gamma$ obtained by emulating the measured channel $(1,3)$, and the estimated mean PER ($\bar{\gamma}'$), maxima ($\gamma'_{\mathrm{max}}$), and minima ($\gamma'_{\mathrm{min}}$) of PERs obtained from emulating 100 simulation runs. The used modulation is 64QAM with a convolutional coding rate of $3/4$. \label{fig:fer_analysis_64qam}}

This shows that our channel model is suitable for thorough testing of vehicular communication hardware and software in urban scenarios that include large vehicles.

\section{Conclusions}
\label{sec:conclusions}
In this paper we have presented the results of the first ever \emph{multi-node} vehicular wireless channel sounding measurement campaign. We use our custom-made AIT multi-node channel sounder \cite{Zelenbaba20_1} in two urban scenarios with a large vehicle obstruction, to simultaneously collect time-variant frequency responses of three vehicular wireless communication channels. We analyze and compare time-variant statistics of the measured links in an overtaking scenario and an intersection scenario to obtain new insights into the impact of large obstructing vehicles on vehicular communication channels.

In the analyzed scenarios the obstructing bus increases the RMS delay spread and RMS Doppler spread by more than $200\,\mathrm{ns}$ and $100\,\mathrm{Hz}$, respectively, compared to the unobstructed link.

The measurement data is used to calibrate a small set of scatterer distribution parameters of the GSCM. Publicly available OpenStreetMap data is used to build the model geometry. We parameterize the path loss caused by the obstructing vehicle and define criteria for modeling obstructing vehicle reflection points, while keeping the added model complexity low. The scenario simulations of the model show an excellent match when compared to the measured links, both qualitatively and quantitatively.

The estimated mean second-order statistics obtained from the simulations show less than $3.6\,\mathrm{dB}$ offset from the measured path loss and $78\,\mathrm{ns}$ and $52\,\mathrm{Hz}$ offset from the measured RMS delay spread and RMS Doppler spread, respectively, for $80\%$ of the samples.

The measured and simulated frequency responses enable us to obtain high resolution time-variant PERs by using a HiL setup with our AIT channel emulator \cite{Hofer19}. We used the obtained PERs to show the consistency of the model quality at the link-level for different modulation parameters. The PER obtained from emulating the measured frequency responses falls within the minima and maxima bounds of PERs obtained from emulating frequency responses of $100$ simulation runs $86\%$ of the time when using QPSK and $98\%$ of the time in the case of 64QAM, showing a consistently excellent match of our model at the link-level.

\appendices
\section{Wireless Channel Characterization}
\label{app:characterization}

The underlying fading process in rapidly time-varying wireless channels, such as the ones in vehicular communications, is non-stationary and its statistical characterization is only valid for a limited stationarity region \cite{Matz05}, \cite{Bernado14}. A stationarity region is a spatial region where we assume the process to be weak stationary and we further assume uncorrelated scattering. We use the local scattering function (LSF) \cite{Matz05} to estimate the scattering function locally for each stationarity region. The LSF is calculated from the recorded frequency responses for each $M\times N$ sample sized stationarity region as a multi-taper estimate \cite{Thomson82} of the scattering function. Since we use time to index measurement data samples, we use $k$ to index the time interval during which a node covers a stationarity region. The LSF is obtained as in \cite{Bernado14}:

\begin{equation}
\hat{C}_{(a,b)}[k;n,p]=\frac{1}{IJ}\sum_{w=0}^{IJ-1}|\mathcal H^{(G_w)}_{(a,b)}[k;n,p]|^2\,,
\label{LSF}
\end{equation}
with the stationarity region length of $T_\mathrm{stat}$, taking $M=T_\mathrm{stat}/T_\mathrm{sys}$ samples in time, and taking all the available $N=Q$ samples in frequency \cite{Bernado12}. The corresponding resolution in the delay domain is thus $\Delta\tau=1/B$ and the resolution in the Doppler domain is $\Delta\nu=1/T_\mathrm{stat}$. To index the relative time sample of each stationarity region we use $m'\in\{ -M/2,...,M/2-1\}$ which relates to the absolute time index as $m=M/2(2k-1)+m'+1$. 

The windowed time-variant frequency response of each link $\mathcal H^{(G_w)}_(a,b)$ is defined as
\begin{multline}
\mathcal H^{(G_w)}_{(a,b)}[k;n,p]=\sum_{m'=-M/2}^{M/2-1}\sum_{q=0}^{N-1}\\ g_{(a,b)}[m'-k,q-k]G_w[m',q]e^{-\mathrm{j}2\pi(pm'-nq)} \,,
\end{multline}
with $n\in\{ 0,...,N-1\}$ and $p\in\{ -M/2,...,M/2-1\}$ denoting discrete delay and Doppler shift, respectively. The window function $G_w[m',q]=u_i[m'+M/2]\tilde{u}_j[q+N/2]$ uses band-limited discrete prolate spheroidal (DPS) sequences \cite{Slepian78} $u_i$, indexed by $i\in\{ 0,...,I-1\}$, and $u_j$, indexed by $j\in\{ 0,...,J-1\}$, while $w=iJ+j$ and $I=J=3$ as in \cite{Bernado09}. 

The time-variant PDP and DSD are used to describe the time-variant second-order statistics of the channel. They are given as the marginals with respect to Doppler 
\begin{equation}
\hat{P}_{\tau; \,(a,b)}[k;n]=\frac{1}{M}\sum_{p=-M/2}^{M/2-1}\hat{C}_{(a,b)}[k;n,p]\,,
\label{PDP}
\end{equation}
and delay
\begin{equation}
\hat{P}_{\nu; \,(a,b)}[k;p]=\frac{1}{N}\sum_{n=0}^{N-1}\hat{C}_{(a,b)}[k;n,p]\,.
\label{DSD}
\end{equation}

The RMS delay spread and RMS Doppler spread are used to describe the delay and Doppler dispersion, of the wireless channel for stationarity region $k$. The RMS delay spread is computed as the second central moment
\begin{equation} 
	\sigma_{\tau; \,(a,b)}[k]=\sqrt{
	\frac{\sum\limits_{n=0}^{N-1}(n\tau_\mathrm{s})^2\hat{P}_{\tau; \,(a,b)}[k;n]}
	{\sum\limits_{n=0}^{N-1}\hat{P}_{\tau; \,(a,b)}[k;n]} -\bar{\tau}_{(a,b)}[k]^2 }\,,
	\label{eq:rms_delay}
\end{equation}
where
\begin{equation} 
    \bar{\tau}_{(a,b)}[k]=\frac{\sum\limits_{n=0}^{N-1}(n\tau_\mathrm{s})\hat{P}_{\tau; \,(a,b)}[k;n]}
    {\sum\limits_{n=0}^{N-1}\hat{P}_{\tau; \,(a,b)}[k;n]} \,\,, 
\end{equation}
is the mean delay. 

The RMS Doppler spread is calculated as
\begin{equation} 
	\sigma_{\nu; \,(a,b)}[k]=\sqrt{
	\frac{\sum\limits_{p=-M/2}^{M/2-1}(p\nu_\mathrm{s})^2\hat{P}_{\nu; \,(a,b)}[k;p]}  
	{\sum\limits_{\mathit{p}=-M/2}^{M/2-1}\hat{P}_{\nu; \,(a,b)}[k;p]} - \bar{\nu}_{(a,b)}[k]^2   }\,,
	\label{eq:rms_Doppler}
\end{equation}
where
\begin{equation}
	\bar{\nu}_{(a,b)}[k]=\frac{\sum\limits_{p=-M/2}^{M/2-1}(p\nu_\mathrm{s})\hat{P}_{\nu; \,(a,b)}[k;p]}{\sum\limits_{\mathit{p}=-M/2}^{M/2-1}\hat{P}_{\nu; \,(a,b)}[k;p]}\,, 
\end{equation}
is the mean Doppler shift.

In calculating (\ref{eq:rms_delay}) and (\ref{eq:rms_Doppler}) we only consider the components of the PDP and the DSD that fulfill a power threshold criterion \cite{Czink07}. This way we omit spurious noise components up to $5\,\mathrm{dB}$ above the noise floor and weak components that are more than $40\,\mathrm{dB}$ below the instantaneous peak value due to receiver sensitivity limitations.

\section{Interpolation of Channel Measurement Data}
\label{app:interpolation}

Channel sounding measurement data is typically acquired with a sampling time $T_\text{s}$, chosen according to the maximum Doppler bandwidth. For channel emulation the convolution of the input signal with the channel impulse response needs to be computed, where the channel impulse response must be provided with sampling time $T_\text{e}=1/B_\text{e}$ defined by the system bandwidth $B_\text{e}$. Typically, $T_\text{s} \gg T_\text{e}$ while the sampling distance in frequency for the measurements $F_\text{s}$ and emulation $F_\text{e}$ are in a similar range. 

Hence, in this appendix we describe an efficient interpolation approach for the above described problem setting. We assume that the parameters $T_\text{s}$, $T_\text{e}$, $F_\text{s}$, and $F_\text{e}$ are appropriately scaled integer numbers. For the interpolation we need to find suitable intermediate sample spacings in time $T_\text{i}$ and frequency $F_\text{i}$ that are largest common divisors, fulfilling
\begin{align}		
T_\text{s}/T_\text{i} =r_\text{t,s} \,, T_\text{e}/T_\text{i} =r_\text{t,e} 
\label{eq:gcd_time}
\end{align}
and
\begin{align}
F_\text{s}/F_\text{i} =r_\text{f,s} \,, F_\text{e}/F_\text{i} =r_\text{f,e} 
\label{eq:gcd_freq}
\end{align}
with $\{r_\text{t,s}, r_\text{t,e}, r_\text{f,s}, r_\text{f,e}\} \in \mathbb{N}$ .

We want to obtain an estimate of the sampled frequency responses on the interpolation time frequency grid\be
g[m,q] = g(m T_\text{i},\varphi(q) F_\text{i})
\label{eq:sampledCFR}
\ee
from subsampled noisy measurements obtained with sampling time $T_\text{s}$ and with sampling distance in frequency $F_\text{s}$. Here, $g(t,f)$ denotes the time-variant channel frequency of the physical propagation channel including the effects of the transmitter and receiver filters, $m$ is the discrete time index and $q$ the discrete frequency index of the interpolated channel frequency response. The function $\varphi(q)=((q+N_\text{i}/2 \mod N_\text{i}) - N_\text{i}/2)$ maps the subcarrier index $q\in\{0,\ldots, N_\text{i}-1\}$ onto the discrete frequency index $\varphi(q)  \in \{-N_\text{i}/2, \ldots, 0, \ldots, N_\text{i}/2-1\}$.

Equation \eqref{eq:sampledCFR} is defined for the region $\mathcal{I}$ in time and frequency, given by the Cartesian product
\begin{equation}
	\mathcal{I}=I^\text{t} \times I^\text{f} = [0,\ldots,M_\text{i}-1]\times [0,\ldots,N_\text{i}-1],
\end{equation}
where $M_\text{i}=(M_\text{s}+2\Delta)r_\text{t,s}$ denotes the number of samples in time and $N_\text{i}=N_\text{s}r_\text{f,s}$ the number of samples in frequency after interpolation. The number of measurement samples in time and frequency are denoted by $M_\text{s}$ and $N_\text{s}$, respectively. We overlap the time frequency regions by $\Delta$ samples to compensate for the increased estimation variance at the block boundaries, see \cite[(32)]{Zemen05b}. 

The noisy measurements, obtained by channel sounding, are indexed by $m'$ in time and $q'$ in frequency,
\begin{align}
	y[m' r_\text{t,s},q' r_\text{f,s}] & = g(m' T_\text{s},q' F_\text{s}) + z(m'T_\text{s},q'F_\text{s})\nonumber\\
	&= g[m'r_\text{t,s},q' r_\text{f,s}] + z[m' r_\text{t,s},q' r_\text{f,s}],
\end{align}
with $T_\text{s}= T_\text{i} r_\text{t,s}$ sample spacing in time, $F_\text{s}= F_\text{i} r_\text{f,s}$ sample spacing in frequency, $m'\in\{0,\ldots, M_\text{s}+2\Delta\}$ and $q'\in\{0,\ldots, N_\text{s}+2\Delta\}$. Additive complex white noise is denoted by $z[m,q]$ with $\mathbb{E}[z] = 0$ and $\text{Var}[z] = \sigma^2$.

For the interpolation we assume a maximum relative velocity $v_{\text{max}}$ and a maximum time delay $\tau_\text{max}$. It follows that the frequency response $g[m,q]$ is band-limited to the region
\begin{equation}
	\mathcal{W} = W^\text{t} \times W^\text{f} = [-\nu_\text{max},\nu_\text{max}] \times [0,\theta_\text{max}],
\end{equation}
with the one sided maximum normalized Doppler bandwidth $\nu_\text{max}=T_\text{i}f_\text{c}v_\text{max}/c_0$, where $f_\text{c}$ is the carrier frequency and $c_0$ denotes the speed of light. The maximum normalized delay $\theta_\text{max}=F_\text{i} \tau_\text{max}$. 

With these assumptions $g[m,q]$ can be estimated using a projection on a two-dimensional DPS subspace as explained in \cite{Zemen12, Zemen12a} for the case of pilot based channel estimation. We will adapt the notation of \cite{Zemen12} for the interpolation problem in this appendix.

Modifying \cite[(20)]{Zemen12} we obtain 
\begin{multline}
y[m,q]={\Big(}\sum_{\ell=0}^{D_t-1}\sum_{k=0}^{D_f-1} u_\ell[m,\mathcal{W}_t,M_\text{i}]\\ \cdot u_k[\varphi(q)+\frac{N_\text{i}}{2},\mathcal{W}_f,N_\text{i}]\psi_{\ell,k}\Big{)}+z[m,q]\,.
\label{eq:SignalModelSubspace}
\end{multline}
where $u_\ell[m,\mathcal{W},M]$ denotes the generalized DPS sequences time-limited to $m\in\{0,\ldots, M \}$ and band-limited to the region $\mathcal{W}$. Please see \cite[(14)-(17)]{Zemen12} for the detailed definition of generalized DPS sequences. The DPS coefficients are denoted by $\psi_{\ell,k}$, the time domain subspace dimension is denoted by $D_t$ and the frequency domain subspace dimension by $D_f$, respectively.

For the purpose of estimating the DPS coefficients $\psi_{i,k}$ we rewrite \eqref{eq:SignalModelSubspace} in matrix vector notation as follows. We collect the coefficient $\psi_{i,k}$ in the vector
\be
\Vec{\psi}=[\Vec{\psi}_0\oT, \ldots, \Vec{\psi}_{D_t-1}\oT ]\oT\, \in \mathbb{C}^{D_tD_f},
\ee
where
\be
\Vec{\psi}_\ell=[\psi_{\ell,0},\ldots, \psi_{\ell,D_f-1}]\oT\,.
\ee

We define the observation vector 
\begin{multline}
\Vec{y}=\left[ y[0,0], \ldots, y[0,N_\text{i}-1], \ldots, \right.\\ 
\left. y[M_\text{i}-1,0],\ldots, y[M_\text{i}-1,N_\text{i}-1]\right]\oT\, \in \mathbb{C}^{M_\text{i} N_\text{i}},
\label{eq:StackedReceiveVector}
\end{multline}
and similarly we define vector $\Vec{g}$ containing the channel frequency response samples $g[m,q]$ and the noise vector $\Vec{z}$ containing the noise values $z[m,q]$.

We define the vector
\be
\Vec{f}[m,\mathcal{W}_t,M_\text{i}]=[u_0[m,\mathcal{W}_t,M_\text{i}],\ldots,u_{D_t-1}[m,\mathcal{W}_t,M_\text{i}]]\oT
\ee
containing the elements of the generalized DPS basis sequences for a given time index $m$. Finally, we define the $M_\text{i} N_\text{i}\times D_tD_f$ matrix
\be\footnotesize
\Vec{\mathcal{D}}=\left[
\begin{array}{c} 
\Vec{f}[0,\mathcal{W}_t,M_\text{i}]\oT \otimes \Vec{f}[\varphi(0)+\frac{N_\text{i}}{2},\mathcal{W}_f,N_\text{i}]\oT \\
\vdots  \\
\Vec{f}[M_\text{i}-1,\mathcal{W}_t,M_\text{i}]\oT \otimes \Vec{f}[\varphi(0)+\frac{N_\text{i}}{2},\mathcal{W}_f,N_\text{i}]\oT \\
\vdots \\
\vdots \\
\Vec{f}[0,\mathcal{W}_t,M_\text{i}]\oT \otimes \Vec{f}[\varphi(N_\text{i}-1)+\frac{N_\text{i}}{2},\mathcal{W}_f,N_\text{i}]\oT \\
\vdots \\
\Vec{f}[M_\text{i}-1,\mathcal{W}_t,M_\text{i}]\oT \otimes \Vec{f}[\varphi(N_\text{i}-1)+\frac{N_\text{i}}{2},\mathcal{W}_f,N_\text{i}]\oT
\end{array}\right]
\label{eq:calD}
\ee
allowing to write the signal model for estimating the generalized DPS coefficient vector $\Vec{\psi}$ as
\be
\Vec{y}=\Vec{\mathcal{D}}\Vec{\psi}+\Vec{z},
\ee
and their least square estimate as
\be
\Vec{\hat{\psi}}=\underbrace{\left(\Vec{\mathcal{D}}\oH \diag(\Vec{d})\Vec{\mathcal{D}}\right)^{-1}\Vec{\mathcal{D}}\oH}_{\Vec{A}\oH}\Vec{y}= \Vec{A}\oH\Vec{y},
\label{eq:LSRRfilter}
\ee
where $\Vec{d}= \Vec{\alpha} \otimes \Vec{\beta}$ is an indicator vector for the sampling grid with elements
$\alpha_k=\gamma(k,r_\text{t,s})$ and $\beta_k=\gamma(k,r_\text{f,s})$ where
\be
\gamma(k,r)= \left\{
\begin{array}{ll}
1 & \text{for} \quad k \mod r =0; \\
0 & \text{otherwise}.
\end{array}
\right.
\ee
Please note that the matrix inversion in \eqref{eq:LSRRfilter} is only of dimension $D_t D_f\times D_t D_f$ and $\Vec{A}$ can be precomputed. For reconstruction on the emulator time frequency grid defined by $T_\text{e}$ and $F_\text{e}$ we use
\be
g[m'',q''] = \left[\Vec{f}[m''r_{t,e},\mathcal{W}_t,M_\text{i}]\oT \otimes \Vec{f}[\varphi(q''r_{f,e}),\mathcal{W}_f,N_\text{i}] \oT \right] \Vec{\psi} \,,
\ee
with $q''\in\{0,\ldots, N_\text{e}\}$ and $m''\in\{\Delta r_\text{t,s}, \ldots, (M+\Delta)r_\text{t,s}-1\}$.

For the numerical implementation we use the parameters depicted in Table \ref{tab:numparam}.

\begin{table}
\begin{center}
\caption{Numerical implementation parameters}
\label{tab:numparam}
\begin{tabular}{lll} 
\toprule
Name 										& Variable 			   		& Value \\
\midrule
carrier frequency				& $f_\text{c}$				&   $5.9\,\text{GHz}$\\
maximum time delay						& $\tau_\text{max}$		&		$4\,\mu$s\\
maximum velocity				& $v_\text{max}$			&		$100\,\text{km/h}= 27.8\,\text{m/s}$\\
\emph{channel sounding} \\
\quad bandwidth			&	$B_\text{s}$				&   $150.250\,\text{MHz}$\\
\quad time spacing & $T_\text{s}$				&		$500\,\mu$s\\
\quad frequency spacing & $F_\text{s}$				&		$B_\text{s}/N_\text{s}=250\text{kHz}$\\
\quad subcarrier    & $N_\text{s}$        &   $601$ \\
\quad block length  & $M_\text{s}$        &   $64$\\
\emph{emulation} \\
\quad bandwidth \\ 
\quad (two times oversampling) 		&	$B_\text{e}$				&   $20\,\text{MHz}$\\
\quad time spacing& $T_\text{e}$        &   $1/B_\text{e} = 50\, \text{ns}$\\
\quad frequency spacing & $F_\text{e}$				&		$B_\text{e}/N_\text{e}=156.25\text{kHz}$\\
\quad subcarrier    & $N_\text{e}$        &   $128$ \\
\quad block length  & $M_\text{e}$        &   $M_\text{s} r_\text{t,s} = 64000$ \\

\emph{oversampling in time} \\
\quad sounding						& $r_\text{t,s}$				&   $10000$ \\
\quad emulation						& $r_\text{t,e}$				&   $1$ \\

\emph{oversampling in frequency} \\
\quad sounding						& $r_\text{f,s}	$			&   $8$ \\
\quad emulation						& $r_\text{f,e}$				&   $5$ \\

\emph{interpolation} \\
\quad time subspace dim.						& $D_\text{t}$				&   $44$ \\
\quad frequency subspace dim.			& $D_\text{f}$				&   $600$\\
\quad overlap       & $\Delta$            &   $4$\\
\quad subcarrier    & $N_\text{i}$        &   $N_\text{s}r_\text{f,s}=4800$ \\
\quad block length  & $M_\text{i}$        &   $(M_\text{s}+2\Delta)r_\text{t,s}= 72000$ \\ 

\bottomrule
\end{tabular}
\end{center}
\end{table}

\bibliography{IEEEabrv, Zelenbaba_master_library}

\begin{thebibliography}{10}
\providecommand{\url}[1]{#1}
\csname url@samestyle\endcsname
\providecommand{\newblock}{\relax}
\providecommand{\bibinfo}[2]{#2}
\providecommand{\BIBentrySTDinterwordspacing}{\spaceskip=0pt\relax}
\providecommand{\BIBentryALTinterwordstretchfactor}{4}
\providecommand{\BIBentryALTinterwordspacing}{\spaceskip=\fontdimen2\font plus
\BIBentryALTinterwordstretchfactor\fontdimen3\font minus
  \fontdimen4\font\relax}
\providecommand{\BIBforeignlanguage}[2]{{%
\expandafter\ifx\csname l@#1\endcsname\relax
\typeout{** WARNING: IEEEtran.bst: No hyphenation pattern has been}%
\typeout{** loaded for the language `#1'. Using the pattern for}%
\typeout{** the default language instead.}%
\else
\language=\csname l@#1\endcsname
\fi
#2}}
\providecommand{\BIBdecl}{\relax}
\BIBdecl

\bibitem{Chen2017}
S.~{Chen}, J.~{Hu}, Y.~{Shi}, Y.~{Peng}, J.~{Fang}, R.~{Zhao}, and L.~{Zhao},
  ``Vehicle-to-everything {(V2X)} services supported by {LTE}-based systems and
  {5G},'' \emph{IEEE Communications Standards Magazine}, vol.~1, no.~2, pp.
  70--76, 2017.

\bibitem{5g2016case}
{5G Automotive Association (5GAA) }, ``The case for cellular {V2X} for safety
  and cooperative driving,'' \emph{White Paper}, Nov. 2016.

\bibitem{Eurostat_transport2020}
Eurostat, ``Passenger transport statistics,'' \emph{Statistics Explained}, vol.
  https://ec.europa.eu/eurostat/statistics-explained/pdfscache/1132.pdf, Jul.
  2020.

\bibitem{Almers_07_channelmodels}
P.~Almers, E.~Bonek, A.~Burr, N.~Czink, M.~Debbah, V.~Degli-Esposti,
  H.~Hofstetter, P.~Ky{\"o}sti, D.~Laurenson, G.~Matz \emph{et~al.}, ``Survey
  of channel and radio propagation models for wireless {MIMO} systems,''
  \emph{EURASIP Journal on Wireless Communications and Networking}, pp. 1--19,
  2007.

\bibitem{OSM}
``{OpenStreetMap.}'' [Online]. Available: https://www.openstreetmap.org.

\bibitem{Wassie19}
D.~A. {Wassie}, I.~{Rodriguez}, G.~{Berardinelli}, F.~M.~L. {Tavares}, T.~B.
  {Sørensen}, T.~L. {Hansen}, and P.~{Mogensen}, ``An agile multi-node
  multi-antenna wireless channel sounding system,'' \emph{IEEE Access}, vol.~7,
  pp. 17\,503--17\,516, 2019.

\bibitem{Almers08}
P.~Almers, K.~Haneda, J.~Koivunen, V.-M. Kolmonen, A.~Molisch, A.~Richter,
  J.~Salmi, F.~Tufvesson, and P.~Vainikainen, ``A dynamic multi-link {MIMO}
  measurement system for 5.3 {GHz},'' \emph{in Proc. 29th URSI General
  Assembly, Chicago, USA}, Aug. 2008.

\bibitem{Bauch07}
G.~{Bauch}, J.~B. {Andersen}, C.~{Guthy}, M.~{Herdin}, J.~{Nielsen}, J.~A.
  {Nossek}, P.~{Tejera}, and W.~{Utschick}, ``Multiuser {MIMO} channel
  measurements and performance in a large office environment,'' in \emph{IEEE
  Wireless Communications and Networking Conference (WCNC), Hong Kong, China},
  2007, pp. 1900--1905.

\bibitem{Bui13}
H.~P. {Bui}, Y.~{Ogawa}, T.~{Nishimura}, and T.~{Ohgane}, ``Performance
  evaluation of a multi-user {MIMO} system with prediction of time-varying
  indoor channels,'' \emph{IEEE Transactions on Antennas and Propagation},
  vol.~61, no.~1, pp. 371--379, 2013.

\bibitem{Zelenbaba20_1}
S.~Zelenbaba, D.~Löschenbrand, M.~Hofer, A.~Dakić, B.~Rainer, G.~Humer, and
  T.~Zemen, ``A scalable mobile multi-node channel sounder,'' in \emph{IEEE
  Wireless Communications \& Networking Conference (WCNC), Seoul, South Korea},
  May 2020.

\bibitem{Nilsson15}
M.~G. {Nilsson}, D.~{Vlastaras}, T.~{Abbas}, B.~{Bergqvist}, and
  F.~{Tufvesson}, ``On multilink shadowing effects in measured {V2V} channels
  on highway,'' in \emph{European Conference on Antennas and Propagation
  (EuCAP), Lisbon, Portugal}, Apr. 2015.

\bibitem{Nilsson18}
M.~G. Nilsson, C.~Gustafson, T.~Abbas, and F.~Tufvesson, ``A path loss and
  shadowing model for multilink vehicle-to-vehicle channels in urban
  intersections,'' \emph{Sensors, vol. 12, no. 4433.}, Dec. 2018.

\bibitem{Gallagher06}
B.~{Gallagher}, H.~{Akalsuka}, and H.~{Suzuki}, ``Wireless communications for
  vehicle safety: Radio link performance and wireless connectivity methods,''
  \emph{IEEE Vehicular Technology Magazine}, vol.~1, no.~4, pp. 4--24, Dec
  2006.

\bibitem{He14}
R.~{He}, A.~F. {Molisch}, F.~{Tufvesson}, Z.~{Zhong}, B.~{Ai}, and T.~{Zhang},
  ``Vehicle-to-vehicle propagation models with large vehicle obstructions,''
  \emph{IEEE Transactions on Intelligent Transportation Systems}, vol.~15,
  no.~5, pp. 2237--2248, Oct 2014.

\bibitem{Vlastaras14}
D.~{Vlastaras}, T.~{Abbas}, M.~{Nilsson}, R.~{Whiton}, M.~{Olbäck}, and
  F.~{Tufvesson}, ``Impact of a truck as an obstacle on vehicle-to-vehicle
  communications in rural and highway scenarios,'' in \emph{IEEE International
  Symposium on Wireless Vehicular Communications (WiVeC), Vancouver, BC,
  Canada}, Sep. 2014.

\bibitem{Mahler16}
K.~{Mahler}, W.~{Keusgen}, F.~{Tufvesson}, T.~{Zemen}, and G.~{Caire},
  ``Propagation channel in a rural overtaking scenario with large obstructing
  vehicles,'' in \emph{2016 IEEE 83rd Vehicular Technology Conference (VTC
  Spring), Nanjing China}, May 2016.

\bibitem{Yang_20}
M.~{Yang}, B.~{Ai}, R.~{He}, G.~{Wang}, L.~{Chen}, X.~{Li}, C.~{Huang},
  Z.~{Ma}, Z.~{Zhong}, J.~{Wang}, Y.~{Li}, and T.~{Juhana}, ``Measurements and
  cluster-based modeling of vehicle-to-vehicle channels with large vehicle
  obstructions,'' \emph{IEEE Transactions on Wireless Communications}, vol.~19,
  no.~9, pp. 5860--5874, 2020.

\bibitem{Vlastaras17}
D.~{Vlastaras}, R.~{Whiton}, and F.~{Tufvesson}, ``A model for power
  contributions from diffraction around a truck in vehicle-to-vehicle
  communications,'' in \emph{2017 15th International Conference on ITS
  Telecommunications (ITST), Warsaw, Poland}, 2017.

\bibitem{Karedal09}
J.~{Karedal}, F.~{Tufvesson}, N.~{Czink}, A.~{Paier}, C.~{Dumard}, T.~{Zemen},
  C.~F. {Mecklenbrauker}, and A.~F. {Molisch}, ``A geometry-based stochastic
  {MIMO} model for vehicle-to-vehicle communications,'' \emph{IEEE Transactions
  on Wireless Communications}, vol.~8, no.~7, Jul. 2009.

\bibitem{Gustafson20}
C.~Gustafson, K.~Mahler, D.~Bolin, and F.~Tufvesson, ``The {COST IRACON}
  geometry-based stochastic channel model for vehicle-to-vehicle communication
  in intersections,'' \emph{IEEE Transactions on Vehicular Technology},
  vol.~69, no.~3, pp. 2365--2375, Mar. 2020.

\bibitem{USRP}
``{USRP-2954 Specifications - National Instruments},'' [Online]. Available:
  https://www.ni.com/pdf/manuals/375725c.pdf.

\bibitem{Friese97}
M.~{Friese}, ``Multitone signals with low crest factor,'' \emph{IEEE
  Transactions on Communications}, vol.~45, no.~10, pp. 1338--1344, Oct 1997.

\bibitem{Molisch10}
A.~F. Molisch, \emph{Wireless Communications}.\hskip 1em plus 0.5em minus
  0.4em\relax 2nd ed., John Wiley \& Sons, 2011.

\bibitem{Zelenbaba19_2}
S.~Zelenbaba, M.~Hofer, D.~Löschenbrand, G.~Kail, M.~Schiefer, and T.~Zemen,
  ``Spatial properties of industrial wireless ultra-reliable low-latency
  communication {MIMO} links,'' in \emph{Asilomar Conference on Signals,
  Systems, and Computers, Pacific Grove (CA), USA}, Nov. 2019.

\bibitem{Paier08}
A.~{Paier}, T.~{Zemen}, L.~{Bernadó}, G.~{Matz}, J.~{Karedal}, N.~{Czink},
  C.~{Dumard}, F.~{Tufvesson}, A.~F. {Molisch}, and C.~F. {Mecklenbrauker},
  ``Non-wssus vehicular channel characterization in highway and urban scenarios
  at 5.2ghz using the local scattering function,'' in \emph{International ITG
  Workshop on Smart Antennas, Darmstadt, Germany}, Feb. 2008, pp. 9--15.

\bibitem{Dakic_21}
A.~Dakić, M.~Hofer, B.~Rainer, S.~Zelenbaba, L.~Bernadó, and T.~Zemen,
  ``Real-time vehicular wireless system-level simulation,'' \emph{IEEE Access},
  vol.~9, pp. 23\,202--23\,217, 2021.

\bibitem{Rain2004:Optimized}
B.~Rainer, M.~Hofer, L.~Bernadó, D.~{L{\"o}schenbrand}, S.~Zelenbaba,
  A.~Dakić, and T.~Zemen, ``Optimized diffuse scattering selection for large
  area real-time geometry-based stochastic modeling of vehicular communication
  links,'' in \emph{IEEE MTT-S International Conference on Microwaves for
  Intelligent Mobility (ICMIM), Linz, Austria}, Dec. 2020.

\bibitem{Adegoke2016VegetationAA}
A.~S. Adegoke, D.~A.~T. Siddle, and S.~O. Salami, ``Vegetation attenuation and
  its dependence on foliage density,'' \emph{European Journal of Engineering
  and Technology}, vol.~4, no.~3, 2016.

\bibitem{Shu18}
X.~{Shu}, C.~{Li}, W.~{Chen}, J.~{Yu}, and K.~{Yang}, ``Performance analysis of
  {V2V} radio channel under typical urban intersection scenario,'' in
  \emph{IEEE International Conference on Communication Systems (ICCS), Chengdu,
  China}, Dec 2018, pp. 216--220.

\bibitem{Zelenbaba20_2}
S.~Zelenbaba, B.~Rainer, M.~Hofer, A.~Dakić, D.~Löschenbrand, and T.~Zemen,
  ``Packet error rate based validation method for an {OpenStreetMap}
  geometry-based channel model,'' in \emph{IEEE 92nd Vehicular Technology
  Conference (VTC-Fall), Victoria, BC, Canada}, Nov. 2020.

\bibitem{Hofer19}
M.~{Hofer}, Z.~{Xu}, D.~{Vlastaras}, B.~{Schrenk}, D.~{Löschenbrand},
  F.~{Tufvesson}, and T.~{Zemen}, ``Real-time geometry-based wireless channel
  emulation,'' \emph{IEEE Transactions on Vehicular Technology}, vol.~68,
  no.~2, pp. 1631--1645, Feb. 2019.

\bibitem{80211p}
IEEE, ``{IEEE Standard for Information technology-- Local and metropolitan area
  networks-- Specific requirements-- Part 11: Wireless LAN Medium Access
  Control (MAC) and Physical Layer (PHY) Specifications Amendment 6: Wireless
  Access in Vehicular Environments},'' \emph{IEEE Std 802.11p-2010 (Amendment
  to IEEE Std 802.11-2007 as amended by IEEE Std 802.11k-2008, IEEE Std
  802.11r-2008, IEEE Std 802.11y-2008, IEEE Std 802.11n-2009, and IEEE Std
  802.11w-2009)}, 2010.

\bibitem{Cohda}
\BIBentryALTinterwordspacing
{Cohda Wireless MK5 OBU specifications}. [Online]. Available:
  \url{https://cohdawireless.com/solutions/hardware/mk5-obu/}
\BIBentrySTDinterwordspacing

\bibitem{Slepian78}
D.~{Slepian}, ``Prolate spheroidal wave functions, fourier analysis, and
  uncertainty — {V}: the discrete case,'' \emph{The Bell System Technical
  Journal}, vol.~57, no.~5, pp. 1371--1430, May 1978.

\bibitem{Matz05}
G.~{Matz}, ``On non-{WSSUS} wireless fading channels,'' \emph{IEEE Transactions
  on Wireless Communications}, vol.~4, no.~5, pp. 2465--2478, Sep. 2005.

\bibitem{Bernado14}
L.~{Bernad{\'o}}, T.~{Zemen}, F.~{Tufvesson}, A.~F. {Molisch}, and C.~F.
  {Mecklenbr{\"a}uker}, ``Delay and {Doppler} spreads of nonstationary
  vehicular channels for safety-relevant scenarios,'' \emph{IEEE Transactions
  on Vehicular Technology}, vol.~63, no.~1, pp. 82--93, Jan. 2014.

\bibitem{Thomson82}
D.~J. {Thomson}, ``Spectrum estimation and harmonic analysis,''
  \emph{Proceedings of the IEEE}, vol.~70, no.~9, pp. 1055--1096, 1982.

\bibitem{Bernado12}
L.~{Bernadó}, T.~{Zemen}, F.~{Tufvesson}, A.~F. {Molisch}, and C.~F.
  {Mecklenbräuker}, ``The (in-) validity of the {WSSUS} assumption in
  vehicular radio channels,'' in \emph{IEEE International Symposium on
  Personal, Indoor and Mobile Radio Communications (PIMRC), Sydney, NSW,
  Australia}, Sep. 2012, pp. 1757--1762.

\bibitem{Bernado09}
L.~{Bernadó}, T.~{Zemen}, A.~{Paier}, J.~{Karedal}, and B.~H. {Fleury},
  ``Parametrization of the local scattering function estimator for
  vehicular-to-vehicular channels,'' in \emph{IEEE Vehicular Technology
  Conference (VTC-Fall), Anchorage, AK, USA}, Sep. 2009.

\bibitem{Czink07}
N.~Czink, ``The random-cluster model -- a stochastic {MIMO} channel model for
  broadband wireless communication systemsof the 3rd generation and beyond,''
  Ph.D. dissertation, TU Wien, Vienna, 2007.

\bibitem{Zemen05b}
T.~Zemen and C.~F. Mecklenbr{\"a}uker, ``Time-variant channel estimation using
  discrete prolate spheroidal sequences,'' \emph{{IEEE} Trans. Signal
  Process.}, vol.~53, no.~9, pp. 3597--3607, September 2005.

\bibitem{Zemen12}
T.~Zemen, L.~Bernado, N.~Czink, and A.~F. Molisch, ``Iterative time-variant
  channel estimation for 802.11p using generalized discrete prolate spheroidal
  sequences,'' \emph{{IEEE} Trans. Veh. Technol.}, vol.~61, no.~3, pp.
  1222--1233, March 2012.

\bibitem{Zemen12a}
T.~Zemen and A.~F. Molisch, ``Adaptive reduced-rank estimation of
  non-stationary time-variant channels using subspace selection,'' \emph{{IEEE}
  Trans. Veh. Technol.}, vol.~61, no.~9, pp. 4042--4056, November 2012.

\end{thebibliography}

\begin{IEEEbiography}[{\includegraphics[width=1in,height=1.25in,clip,keepaspectratio]{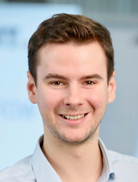}}]{STEFAN ZELENBABA} graduated from the Faculty of Electrical Engineering of University of Belgrade, Telecommunications Engineering in 2015, and received his Master’s degree in 2017 after a collaboration with Nokia Bell Labs in Dublin. Since October 2017 he is a Ph.D. candidate with the Austrian Institute of Technology in the reliable wireless communications research group of Thomas Zemen. His research is focused on measurements and characterization of non-stationary time-variant wireless channels and geometry-based wireless channel modeling.
\end{IEEEbiography} 

\begin{IEEEbiography}[{\includegraphics[width=1in,height=1.25in,clip,keepaspectratio]{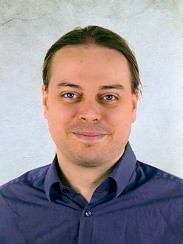}}]{BENJAMIN RAINER} is a researcher in AIT’s Center for Digital Safety and Security. He received the M.Sc. (Dipl.-Ing.), and Ph.D. (Dr. techn.) in computer science all with distinction from the Alpen-Adria-University Klagenfurt. From 2012 to 2016 he has been with the Alpen-Adria-University Klagenfurt as a Researcher and Postdoc at the institute  of Information Technology. He has been with the Austrian Institute of Technology, Vienna since 2017. His research interests are security and communication in future (mobile) networks with emphasises on ultra-reliable low latency wireless communication, wireless channel modelling, real-time simulation and emulation of wireless channels and system-level simulations of communication systems.
\end{IEEEbiography} 

\begin{IEEEbiography}[{\includegraphics[width=1in,height=1.25in,clip,keepaspectratio]{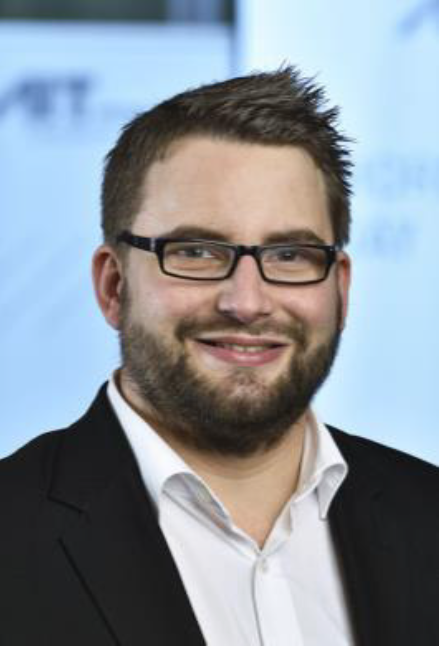}}]{MARKUS HOFER} received the Dipl.-Ing. degree (with distinction) in telecommunications from the Vienna University of Technology, Vienna, Austria, in 2013 and the doctoral degree in 2019. From 2013 to 2015 he was with the FTW Telecommunications Research Center Vienna working as a Researcher in “Signal and Information Processing” department. He has been with the AIT Austrian Institute of Technology, Vienna since 2015 and is working as a Scientist in the research group for ultrareliable wireless machine-to-machine communications.. His research interests include low- latency wireless communications, time-variant channel measurements, modeling and real-time emulation; time-variant channel estimation, 5G massive MIMO systems; software-defined radio rapid prototyping, cooperative communication systems, and interference management. 
\end{IEEEbiography} 

\begin{IEEEbiography}[{\includegraphics[width=1in,height=1.25in,clip,keepaspectratio]{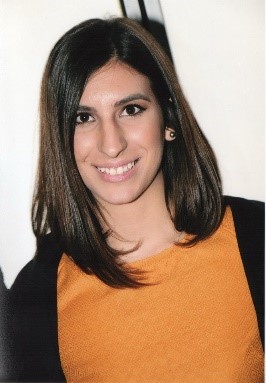}}]{Anja DakiĆ} graduated in the School of Electrical Engineering of University of Belgrade, Telecommunications and Information in 2017, and received the master’s degree in 2019. Since 2019 she is a Ph.D. candidate with the Austrian Institute of Technology in the group of Thomas Zemen. She is a member of the Reliable Wireless Communication Team of the AIT since 2018. Her special areas of research include wireless mobile communication for vehicles, modeling vehicular channels, measurements with hardware modems and system-level simulations of communications systems.
\end{IEEEbiography} \vspace{-12mm}

\begin{IEEEbiography}[{\includegraphics[width=1in,height=1.25in,clip,keepaspectratio]{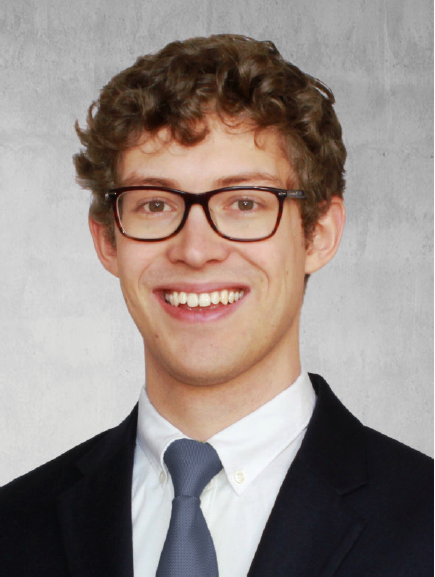}}]{David Löschenbrand} received the Dipl.-Ing. degree (with distinction) in Telecommunications in 2016 from Vienna University of Technology. From 2012 to 2015, he worked for the Institute of Telecommunications, implementing software for antenna characterization purposes. Since 2016, he is a Ph.D. candidate with the Austrian Institute of Technology in the group of Thomas Zemen. His research interests focus on massive MIMO in time-varying propagation channels, channel aging, antenna design, reliable low-latency wireless communications for highly autonomous vehicles, vehicular channel measurements and channel modelling. He gained experience in research work and project management during several finished and ongoing applied research projects on a national and international scale. He successfully implemented a software defined radio based massive MIMO testbed for high-speed measurements and signal processing experimentation, which provides crucial insights for several of AIT’s projects.
\end{IEEEbiography} \vspace{-12mm}

\begin{IEEEbiography}[{\includegraphics[width=1in,height=1.25in,clip,keepaspectratio]{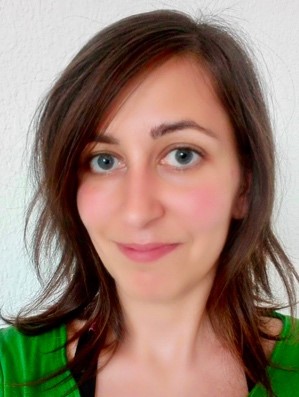}}]{LAURA BERNADÓ}
 obtained her PhD in telecommunications engineering from the Vienna University of Technology (VUT) in 2012 and the M.Sc. degree in telecommunications engineering from the Technical University of Catalonia (UPC) in 2007, with the Master Thesis written at the Royal Institute of Technology (KTH), in Stockholm. Mrs. Bernadó has worked as a researcher in the signal and information processing department at the Telecommunications Research Center in Vienna, Austria, for 6 years, and as an antenna engineer at Fractus SA, Spain, for 3 years. Currently she works as a scientist in the reliable wireless communications research group at the department for digital safety and security at AIT Austrian Institute of Technology. Her research interests are modeling of fast time-varying non-stationary fading processes, channel emulation and transceiver design for ultra-reliable wireless communication systems. 
\end{IEEEbiography} \vspace{-10mm}

\begin{IEEEbiography}[{\includegraphics[width=1in,height=1.25in,clip,keepaspectratio]{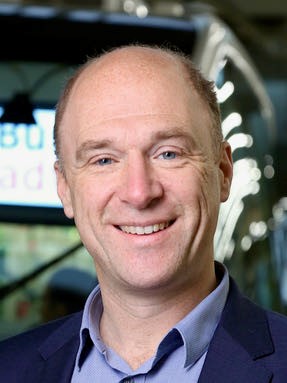}}]{THOMAS ZEMEN }(S’03–M’05–SM’10) received the Dipl.-Ing. degree in electrical engineering in 1998, the doctoral degree in 2004 and the Venia Docendi (Habilitation) in 2013, all from Vienna University of Technology. He is Principal Scientist at the AIT Austrian Institute of Technology, Vienna, Austria, leading the reliable wireless communications group. Previously he worked at the Telecommunication Research Center Vienna (FTW) and Siemens Austria. Mr. Zemen is the author or coauthor of four books chapters, 37 journal papers and more than 113 conference communications. His research interests focus on the interaction of the physical wireless radio communication channel with other parts of a wireless communication system for time-sensitive 5G and 6G machine-to-machine communications. Dr. Zemen is docent at the Vienna University of Technology and served as Editor for the IEEE Transactions on Wireless Communications from 2011 - 2017.

\end{IEEEbiography}

\EOD

\end{document}